\documentclass[preprint,amsmath,amssymb,aps,a4]{revtex4-2}
\usepackage{amsfonts} %\usepackage{citesort}
\usepackage{graphicx} % Include figure files
\usepackage{epsfig}
\usepackage{multirow}
\usepackage{bm}% bold math
\usepackage{color}
\usepackage{amssymb}
\usepackage{dsfont}
\usepackage[colorlinks=true, linkcolor=blue, citecolor=blue, urlcolor=blue]{hyperref}
\usepackage{cleveref}
\usepackage{subfig}
\usepackage{float}

\newcommand{\be}{\begin{eqnarray}}
	\newcommand{\ee}{\end{eqnarray}}

\begin{document}
\raggedbottom
\title{In-medium properties of $D$ and $D^*$ mesons in magnetized isospin asymmetric nuclear matter}
\author{Saksham Sharma}
\email{sakshamsha1234@gmail.com}
\author{Dhananjay Singh}
\email{snaks16aug@gmail.com}
\author{Suneel Dutt}
\email{dutts@nitj.ac.in}
\author{Harleen Dahiya}
\email{dahiyah@nitj.ac.in}
\author{Arvind Kumar}
\email{kumara@nitj.ac.in}
\affiliation{Department of Physics, Dr. B.R. Ambedkar National
	Institute of Technology, Jalandhar, 144008, India}

	\begin{abstract}

    We investigate the impact of an external magnetic field on the in-medium properties of pseudoscalar ($D^0,D^+,D_s$) and vector ($D^{0*},D^{+*},D_{s}^{*}$) mesons in isospin asymmetric nuclear matter at finite temperature using a hybrid theoretical framework combining the chiral SU(3) quark mean-field (CQMF) model and the light-front quark model (LFQM). The medium-modified constituent quark masses, obtained from the CQMF model by including the magnetized Dirac sea contribution and anomalous magnetic moments of nucleons, are used as input to the LFQM calculations of meson masses, weak decay constants, and leading-twist distribution amplitudes. We further incorporate the Landau quantization of the charged mesons restricted to the lowest Landau level, while magnetic field induced pseudoscalar-vector mixing is taken into account for each $D$-$D^{*}$ doublet. We find that the external magnetic field enhances the effective masses and decay constants of both pseudoscalar and vector $D$ mesons by magnetic catalysis, while increasing baryon density generally induces an attractive mass shift and suppresses the decay constant and distribution amplitudes. The Landau level contribution further enhances the effective masses of the charged mesons, whereas the pseudoscalar-vector mixing produces a level repulsion, shifting the vector meson masses upward and the pseudoscalar meson masses downward. The interplay between magnetic field and density effects gives rise to a nontrivial medium behavior of heavy-light meson properties, with isospin asymmetry further inducing a small but systematic mass splitting across all the meson states considered. These results provide useful insights into heavy-flavor dynamics in strongly interacting matter and are relevant to ongoing and future studies at FAIR, NICA, and J-PARC.

	\end{abstract}
	
	\maketitle
	
	\section{\label{intro}Introduction}

Probing the properties of hadrons in hot and dense strongly interacting matter is one of the central challenging topics in hadronic physics as the study of in-medium modifications of hadrons is closely related to observables measured in Heavy-Ion Collision (HIC) experiments \cite{Rapp:1999ej, LI1999619} and plays an important role in understanding the properties of compact astrophysical objects, such as neutron stars and magnetars \cite{Lattimer2021,Turolla_2015,Kaspi2017}. Since hadrons are composite systems governed by the non-perturbative aspects of quantum chromodynamics (QCD), their masses, decay widths, and internal structure undergo significant modifications in a nuclear environment. One of the earliest experimental evidence for in-medium modifications of hadrons was provided by the European Muon Collaboration (EMC), which observed that the structure functions of nucleons bound in nuclei are modified relative to those of free nucleons in deep-inelastic scattering (DIS) experiments \cite{EuropeanMuon:1983wih}. Such medium modifications are linked to fundamental aspects of QCD, in particular to chiral symmetry breaking and its partial restoration at high density and/or temperature. Significant modifications of pion properties in nuclear environment are supported by several experimental observations like low-energy pion-nucleus scattering \cite{FriedmanE.}, deeply bound pionic atoms \cite{Suzuki:2002ae}, and di-pion production in hadron- and photon-nucleus interaction \cite{BonuttiF., CAMERINI200489}. These observations provide strong evidence for the partial restoration of chiral symmetry in a dense nuclear medium. 
\par Several facilities worldwide are dedicated to explore the strongly interacting matter in the non-perturbative regime of QCD, motivating substantial theoretical work on the medium-induced modifications of hadrons. The Japan Proton Accelerator Research Complex (J-PARC) provides high-intensity hadron beams for the study of hadron properties and strangeness physics in dense nuclear matter \cite{Ohnishi_2020}. The PANDA experiment (antiProton ANnihilation at DArmstadt) at the Facility for Antiproton and Ion Research (FAIR) is designed to study hadron spectroscopy and the in-medium masses and widths in the open-charm and bottom sectors \cite{pandacollaboration2009physicsperformancereportpanda}. The Compressed Baryonic Matter (CBM) at GSI, Germany \cite{Senger+2012+1289+1294} and the Nuclotron-based Ion Collider fAcility (NICA) at JINR, Dubna, Russia \cite{article, Kekelidze:2017ghu} aim to probe nonperturbative aspects of QCD at high baryon density. 
\par In addition to finite density and temperature, the influence of magnetic field on hadrons in hot and dense matter has attracted considerable attention in recent years, since strong magnetic field is created in non-central heavy-ion collisions. The magnitude of such magnetic fields have been estimated to be about $eB\sim10-15m_\pi^2$ ($\sim 4\times 10^{19}$ Gauss) at the Large Hadron Collider (LHC), CERN \cite{Skokov:2009qp}, and $eB\sim2m_\pi^2$ at Relativistic Heavy Ion Collider (RHIC), BNL \cite{KHARZEEV2008227}. Strong magnetic fields are associated with dense astrophysical objects, for example surface fields of magnetars reach $\sim10^{14}-10^{15}$ Gauss, while interior fields are expected to be several orders of magnitude higher \cite{Turolla_2015, Kumari:2020mci,PhysRevC.107.035807}. Strong magnetic fields give rise to many interesting physical phenomena, including the enhancement (reduction) of the light-quark condensate with increasing $eB$ known as magnetic catalysis (inverse magnetic catalysis) \cite{Gusynin_1996,Bali:2012zg,Bali}, the chiral magnetic effect (CME) \cite{Kharzeev_2011}, and chiral magnetic wave \cite{Huang_2023}. Apart from modifying the QCD vacuum, the presence of the magnetic field breaks the spatial rotational symmetry of the system and induces a mixing between pseudoscalar ($J^P=0^-$) and the longitudinal component of the corresponding vector ($J^P=1^-$) meson. This pseudoscalar-vector (PV) mixing leads to a level repulsion and modifies the masses of the physical meson states \cite{PhysRevD.91.045025}. For charged mesons, the transverse motion quantizes into discrete Landau levels under the influence of external magnetic field, modifying the single-particle energy spectrum and contributing to the effective masses of these mesons \cite{Gubler:2015qok}. The magnetic fields produced in non-central ultrarelativistic heavy ion collisions decrease rapidly after the collision. However, induced currents in the conducting medium slow this decay and prolong the field's lifetime \cite{PhysRevC.96.034902, PhysRevC.84.044908, PhysRevC.83.039903}. The strong magnetic fields produced in peripheral collisions at the available facilities further motivate theoretical investigations of in-medium hadron properties under the combined effects of finite density, temperature, and external magnetic field. 
\par Various theoretical frameworks have been employed to investigate medium-induced modification of hadron properties, including light-front holographic QCD \cite{PhysRevC.106.055202}, the Nambu-Jona-Lasinio (NJL) model combined with the quark meson coupling (QMC) approach for the valence-quark distributions of pions and kaons in symmetric nuclear matter \cite{PhysRevD.100.094011}, Dyson-Schwinger equation (DSE) based methods \cite{ROBERTS2000S1}, and hybrid light-front quark models (LFQM) \cite{PhysRevD.59.074015}. In particular, the LFQM combined with the QMC model has been applied to explore the distribution amplitudes (DAs) and the weak decay constants of pseudoscalar and vector mesons in symmetric nuclear matter \cite{PhysRevD.107.114010}. The LFQM has also been widely employed to investigate the properties of heavy-light $B$ and $B_s$ mesons, including their masses, decay constants, form factors, and radiative decays in vacuum as well as in-medium environments \cite{PhysRevD.80.054016, PhysRevD.75.073016, Choi:2007us, Tanisha:2025glu}. More recently, the hybrid LFQM combined with the chiral SU(3) quark mean-field (CQMF) model has been employed to investigate the in-medium masses, weak decay constants, and distribution amplitudes of pseudoscalar and vector $D$ mesons in isospin asymmetric nuclear matter \cite{10.1093/ptep/ptaf135}. The presence of a strong magnetic field along with the phenomenon of PV mixing taken into account has been investigated in both the hidden-charm and open-charm sectors, demonstrating significant modifications in the masses of the corresponding mesons \cite{Gubler:2015qok, PhysRevLett.113.172301, PhysRevD.91.045025}. Furthermore, the influence of Landau quantization on the masses of charged heavy mesons has been studied within QCD sum-rule analysis \cite{Mishra:2023uhx,Gubler:2015qok}. Although considerable progress has been made in understanding the medium modifications of heavy mesons, including the effects of PV mixing, Landau quantization, and the baryonic Dirac sea \cite{Mishra:2023uhx}, a simultaneous investigation of the in-medium masses, weak decay constants, and distribution amplitudes of both pseudoscalar and vector $D$ mesons within a hybrid CQMF-LFQM framework, incorporating the baryonic Dirac sea and anomalous magnetic moments (AMMs), has not yet been carried out. In the present work, we compute the in-medium masses, weak decay constants, and distribution amplitudes of the pseudoscalar ($D^0,D^+,D_s$) and vector ($D^{0*},D^{+*},D_s^*$) mesons in magnetized isospin asymmetric nuclear matter using a hybrid CQMF-LFQM framework, incorporating the effects of the baryonic Dirac sea and the AMMs of nucleons.

\par Open heavy flavor mesons, in particular $D$ and $\bar{D}$, have been studied extensively under various conditions of density, temperature, and magnetic field. A large number of non-perturbative  approaches have been used to investigate the effects of strong magnetic field on quark matter, nuclear matter, and compact-star properties, including the QCD sum rules \cite{PhysRevD.93.054035, Machado2014,Parui2022,Mishra:2023uhx}, the QMC model \cite{PhysRevC.77.045804}, relativistic mean field (RMF) models \cite{Broderick_2000, BRODERICK2002167, SINHA201343, PhysRevC.84.035803}, NJL \cite{PhysRevC.79.035807} and its Polyakov-loop extensions (PNJL) \cite{Moreira_2021, PhysRevD.105.074034}, and the chiral hadronic model \cite{Mishra2004}. In Ref.~\cite{PhysRevD.90.125036}, the effects of magnetized Dirac sea on the vacuum to nuclear matter transition were studied using the Walecka model and an extended linear sigma model. The inclusion of the AMMs of nucleons within the Walecka model was shown to lead to inverse magnetic catalysis at finite density and temperature, in contrast to the magnetic catalysis obtained when the AMMs are neglected \cite{PhysRevD.98.056024}. In this study, the AMM contributions were incorporated through a weak-field expansion of the fermionic propagator in evaluating the one-loop nucleon self-energy. Within the chiral effective model framework, the magnetized Dirac sea contribution and the AMM-induced inverse magnetic catalysis at finite baryon density have been incorporated in studies of open and hidden heavy-flavor mesons in nuclear and hyperonic matter \cite{Mishra:2023uhx}. Motivated by these developments, the present work combines the CQMF model with the LFQM to investigate the in-medium properties of the $D$ and $D^{*}$ mesons in magnetized nuclear matter.

\par In the CQMF model, the in-medium masses of mesons are governed by the interactions of the constituent light quarks with the scalar ($\sigma,\zeta,\delta$) and the vector ($\omega,\rho$) meson fields generated by the nuclear medium. In the presence of an external magnetic field, the single particle energy spectrum of charged baryons is modified by Landau quantization and additional contributions arise  from the AMM of all baryons and from the magnetized Dirac sea. The charm quark, being heavy, is only weakly affected by the medium, so the dominant magnetic field induced modifications of $D$ and $D^*$ properties enter through the light constituent quarks $u$, $d$, and $s$. The weak decay constants parameterize the meson to vacuum matrix elements of the axial vector and vector currents and serve as indicators of  chiral symmetry breaking in dense matter \cite{Chhabra:2018oyl, Chhabra_2017}. The leading-twist distribution amplitudes encode the longitudinal momentum distribution of valence quarks within the meson and play a central role in the QCD description of hard exclusive processes \cite{PhysRevD.110.014020}. In the LFQM, hadrons are described as relativistic bound states of constituent quarks quantized on the light-front hypersurface $x^+ = x^0 + x^3$ = const. \cite{PhysRevD.59.034001}. We adopt a Gaussian radial wave function as the variational trial state, in line with standard LFQM practice for heavy-light meson observables. The effects of isospin asymmetry are incorporated through the scalar-isovector $\delta$ field and the vector-isovector $\rho$ field, which acquire non-zero expectation values in isospin asymmetric medium.

\par This paper is organized as follows. In Sec.~\ref{subsec:CQMF}, we describe the CQMF model used to compute the in-medium masses of quarks and Sec.~\ref{subsec:LFQM} describes the LFQM used to calculate the in-medium properties of the pseudoscalar and vector $D$ mesons. The results for the in-medium properties of $D$ and $D^*$ mesons are presented in Sec.~\ref{sec:results}. Section~\ref{sec:Summary} summarizes our findings and outlines directions for future work.

\section{Theoretical Framework}
\label{sec:theory}

\subsection{Chiral SU(3) Quark Mean Field (CQMF) Model}
\label{subsec:CQMF}
In this section, we shall describe the CQMF model used in the present work to study the medium-modified properties of $D$ mesons at finite temperature in the presence of an external magnetic field, for different baryonic densities. 
The effective Lagrangian density of the model can be expressed as \cite{Ping_2001}
\begin{eqnarray}
\label{eq:1}
\mathcal{L}_{\text{eff}} &=& \mathcal{L}_{q0} + \mathcal{L}_{qm} + \mathcal{L}_{\Sigma \Sigma} + \mathcal{L}_{VV} + \mathcal{L}_{\chi SB} +\mathcal{L}_{c} + \mathcal{L}_{\Delta m} + \mathcal{L}_{\text{mag}}.
\end{eqnarray}
Here, ${\cal L}_{q0}$ denotes the kinetic term and ${\cal L}_{qm}$ represents the quark meson interaction term, including quark interaction with both spin-0 and spin-1 mesons. The expressions for ${\cal L}_{q0}$ and ${\cal L}_{qm}$ are respectively given as 
\begin{eqnarray}
\mathcal{L}_{q0} &=& \bar{\Psi} \, i \gamma^\mu \partial_\mu \Psi,
\end{eqnarray}
and 
\begin{eqnarray}
\mathcal{L}_{qm} &=& g_s \left(\bar{\Psi}_L M \Psi_R + \bar{\Psi}_R M^\dagger \Psi_L\right) 
- g_v \left(\bar{\Psi}_L \gamma^\mu l_\mu \Psi_L + \bar{\Psi}_R \gamma^\mu r_\mu \Psi_R\right) \nonumber \\
&=& \frac{g_s}{\sqrt{2}} \, \bar{\Psi} \left( \sum_{a=0}^8 s_a \lambda_a + i \gamma^5 \sum_{a=0}^8 p_a \lambda_a \right) \Psi \nonumber \\
&& - \frac{g_v}{2\sqrt{2}} \, \bar{\Psi} \left( \gamma^\mu \sum_{a=0}^8 v_a^\mu \lambda_a 
- \gamma^\mu \gamma^5 \sum_{a=0}^8 a^\mu_a \lambda_a \right) \Psi.
\end{eqnarray}
 Here, \( \Psi = \begin{pmatrix} u \\ d \\s \end{pmatrix} \) and parameters $g_s$ and $g_v$ represent the coupling of quark with scalar and vector meson fields, respectively. In Eq.~\eqref{eq:1}, the term \(\mathcal{L}_{\Sigma \Sigma}\) represents the self-interaction of scalar mesons and in the mean-field
approximation is expressed as \cite{PhysRevC.59.411}
\begin{equation}
\begin{aligned}
\mathcal{L}_{\Sigma\Sigma} = & -\frac{1}{2} k_0 \chi^2 \left(\sigma^2 + \zeta^2 + \delta^2\right) 
+ k_1 \left(\sigma^2 + \zeta^2 + \delta^2\right)^2 
+ k_2 \left(\frac{\sigma^4}{2} + \frac{\delta^4}{2} + 3\sigma^2\delta^2 + \zeta^4\right) \\
& + k_3 \chi \left(\sigma^2 - \delta^2\right) \zeta 
- k_4 \chi^4 - \frac{1}{4} \chi^4 \ln\frac{\chi^4}{\chi_0^4} 
+ \frac{\xi}{3} \chi^4 \ln \left(\left(\frac{(\sigma^2 - \delta^2)\zeta}{\sigma_0^2\zeta_0}\right) 
\left(\frac{\chi^3}{\chi_{0}^3}\right)\right).
\end{aligned}
\end{equation}
The last two logarithmic terms are used to introduce the scale-breaking effect to compute the trace of energy-momentum tensor within this model. The vector meson self-interaction term, \(\mathcal{L}_{V V}\) is given by
\begin{eqnarray}
\mathcal{L}_{VV} & = & \frac{1}{2} \frac{\chi^2}{\chi_0^2} \left( m^2_\omega \omega^2 + m^2_\rho \rho^2   \right) + g_4 \left( \omega^4  + 6\omega^2\rho^2 + \rho^4  \right).
\end{eqnarray}
For pseudoscalar mesons, non-zero masses are generated by the explicit breaking of chiral symmetry through the Lagrangian density  \cite{PhysRevC.59.411, WANG2002455}
\begin{eqnarray}
\mathcal{L}_{\chi SB} & = &-  \frac{\chi^2}{\chi_0^2} \left[ m_\pi^2 f_\pi \sigma + \left(\sqrt{2} m_K^2 f_K -\frac{ m_\pi^2} {\sqrt{2}} f_\pi \right)\zeta \right].
\end{eqnarray}
To represent quark confinement in baryons, the confining potential term $\mathcal{L}_c = -\bar\psi \chi_c \psi$ is introduced. The Lagrangian density term $\mathcal{L}_{\Delta m}$ in Eq.~\eqref{eq:1} is given by \(\mathcal{L}_{\Delta m}=-(\Delta m)\bar\psi S_1 \psi\) and is incorporated to attain a realistic value for the strange quark mass $m_s$, where $S_1$ represents the strange quark matrix operator and is defined as $S_1=\frac{1}{3}\left(I-\lambda_8 \sqrt{3}\right)=diag(0,0,1)$ \cite{WANG2002455}.
Under the influence of meson mean fields, the Dirac equation for a quark field $\Psi_{qi}$ is expressed as
\begin{eqnarray}
\left[- i \boldsymbol{\alpha} \cdot \nabla + \chi_c(r) + \beta m^*_q \right]\Psi_{qi} = e^*_q \, \Psi_{qi},
\end{eqnarray}
where the subscript $q$ represents the quarks within a baryon of species $i$ (where $i$=$n$, $p$) and $\boldsymbol\alpha$, $\beta$ denote the standard Dirac matrices. The effective quark mass $m_q^*$ is given by the following relation in terms of scalar-isoscalar non-strange ($\sigma$) and strange ($\zeta$) meson fields as \cite{Kumar:2023owb}
\begin{eqnarray}
\label{eq:quark_mass}
m^*_q & = & - g^q_\sigma \sigma - g^q_\zeta \zeta - g_\delta^q I^{3q} \delta,
\end{eqnarray}
where $g_\sigma^q$, $g_\zeta^q$, and $g_\delta^q$ are the coupling constants of quarks to the scalar meson fields $\sigma$, $\zeta$, and $\delta$, respectively. Here, $I^{3q}$ denotes the third component of the isospin associated with quark flavor $q$, with $I^{3u}=\frac{1}{2}$, $I^{3d}=-\frac{1}{2}$, and $I^{3s}=0$. The interaction of constituent quarks with the vector meson fields $\omega$ and $\rho$ modifies the quark chemical potential, such that the effective chemical potential for a quark of flavor $q$ is given by
\begin{eqnarray}
\label{eq:10}
   \mu^*_q &=& \mu_q - g_\omega^q \, \omega - g_\rho^q \, I^{3q} \rho .
\end{eqnarray}   
The effective mass of the baryon $M_i^*$, is related to spurious center of momentum $\langle p^{*2}_{i\:\text{cm}} \rangle$ \cite{PhysRevD.31.1652, PhysRevC.88.015206} and effective quark energy $e_q^*$ as
\begin{eqnarray}
M^*_i & = & \sqrt{\left( \sum_q n_{qi} e^*_q + E_{i\:\text{spin}}\right)^2 - \langle p^{*2}_{i\:\text{cm}} \rangle}.
\end{eqnarray}
Here, $n_{qi}$ denotes the number of quarks of type $q$ within the $i^{th}$ baryon. The spurious center of momentum of baryon $\langle p^{*2}_{i\:\text{cm}} \rangle$ can be expressed in terms of $e_q^*$ and $m_q^*$ via following relation
\begin{eqnarray}
\langle p^{*2}_{i\:\text{cm}}\rangle & = & \sum_q\frac{\left( 11 e^*_q + m^*_q \right)}{ 6\left( 3 e^*_q + m^*_q \right)} \left( e^{*2}_q - m^{*2}_q \right).
\end{eqnarray}
In the presence of an external magnetic field, the interaction of the nucleons with the electromagnetic field is incorporated through the additional Lagrangian density term $\mathcal{L}_{\rm mag}$. Within the present framework, the magnetic field couples directly only to the nucleonic sector via the electromagnetic coupling and the AMM interaction. Corresponding interaction Lagrangian is given as
\begin{eqnarray}
\mathcal{L}_{\text{mag}}=-\bar\Psi_i\ q_i \ \gamma ^\mu A_\mu \Psi \ - \ \frac{1}{2} \kappa_i  \bar\Psi_i \sigma^{\mu\nu}F_{\mu\nu}\Psi_i \ - \frac{1}{4}F^{\mu\nu}F_{\mu\nu},
\end{eqnarray}
where $q_i$ and $\kappa_i$ are the electric charge and AMM of the $i^{th}$ baryon respectively, described by field $\Psi$. The first term describes the standard electromagnetic coupling and the second term represents the interaction of the AMM of the baryon with the electromagnetic field and accounts for the composite internal structure of the baryon beyond the minimal electromagnetic coupling. The electromagnetic field strength tensor is defined as $F_{\mu\nu} = \partial_\mu A_\nu \ -  \partial_\nu A_\mu$, and $\sigma^{\mu\nu}=\frac{i}{2}[\gamma^\mu,\gamma^\nu].$ At finite temperature and density, the thermodynamic potential for magnetized nuclear matter is expressed as
\begin{eqnarray}
\label{eq:14}
    \Omega = \Omega_{\text{DS}} + \Omega_{\text{med}} - \mathcal{L_ {\rm M}}.
\end{eqnarray}
In Eq.~(\ref{eq:14}), $\Omega_{\text{DS}}$ and $\Omega_{\text{med}}$ represent the contribution of Dirac sea and thermal part (Fermi sea), respectively, of the baryons to the thermodynamic potential and $\mathcal{L_ {\rm M}}=\mathcal{L}_ {{\Sigma \Sigma}}+\mathcal{L_ {\rm VV}}+\mathcal{L_ {\chi \rm SB}}$ denotes the interaction between mesons. The quark masses and chemical potentials are modified due to their interaction with the scalar and vector fields. For spin-$\frac{1}{2}$ charged baryons, these Dirac and Fermi sea contributions to the thermodynamic potential are respectively given as \cite{PhysRevD.90.125036,Aguirre:2016vqa,Aguirre:2019ivr} 
\begin{align}
\Omega_{\text{DS}}^{\text{charged}}
= \sum_i \Omega_{\text{DS}}^{i}
= - \sum_i \frac{|q_iB|}{2\pi}
\left[
\sum_{\nu=0}^{\nu_{\max}}
\int_{0}^{\infty}\frac{dk_z}{2\pi}\,
\epsilon_{i,k,\nu,s=+1}
+
\sum_{\nu=1}^{\nu_{\max}}
\int_{0}^{\infty}\frac{dk_z}{2\pi}\,
\epsilon_{i,k,\nu,s=-1}
\right],
\label{ther_pot_Dirac_sea_charged_baryons}
\end{align}
and 
\begin{equation}
\begin{aligned}
\Omega_{\mathrm{med}}^{\mathrm{charged}}
&= \sum_i \Omega_{\mathrm{med}}^{i}
= -T\sum_i \frac{|q_iB|}{2\pi}
\Bigg[
\sum_{\nu=0}^{\nu_{\max}}
\int_{0}^{\infty}\frac{dk_z}{2\pi}
\\
&\qquad\times
\Bigg\{
\ln\!\left(1+e^{-\beta(\epsilon_{i,k,\nu,s=+1}-\mu_i^*)}\right)
+\ln\!\left(1+e^{-\beta(\epsilon_{i,k,\nu,s=+1}+\mu_i^*)}\right)
\Bigg\}
\\
&\qquad+
\sum_{\nu=1}^{\nu_{\max}}
\int_{0}^{\infty}\frac{dk_z}{2\pi}
\Bigg\{
\ln\!\left(1+e^{-\beta(\epsilon_{i,k,\nu,s=-1}-\mu_i^*)}\right)
+\ln\!\left(1+e^{-\beta(\epsilon_{i,k,\nu,s=-1}+\mu_i^*)}\right)
\Bigg\}
\Bigg],
\end{aligned}
\label{ther_pot_Fermi_sea_charged_baryons}
\end{equation}
where the sum over $\nu$ refers to Landau levels. Here, $\beta=1/T$ and $\epsilon_{k,\nu,s}^i$ is the single-fermion energy of the charged baryon given as
\begin{equation}
\epsilon^i_{k,\nu,s} = \sqrt{k_z^2 
+ \left ( \sqrt{2\nu |q_iB|+ {M_i^{*}}^2}-s\kappa_i B\right)^2}.
\end{equation}
For neutral baryon, Dirac sea contribution to thermodynamic potential is expressed as
\cite{PhysRevD.90.125036,Aguirre:2016vqa,Aguirre:2019ivr}
\begin{align}
\Omega_{\text{DS}}^{\text{neutral}} =
\sum _i \Omega_{\text{DS}}^{i} =
 -\sum _i \sum_{s = \pm 1}  \int  
\frac{d^3k}{\left(2\pi\right)^3} \epsilon^i_{k,s},
\label{ther_pot_Dirac_sea_neutral_baryons}
\end{align}
while the Fermi sea contributes as
\begin{equation}
\Omega_{\rm med}^{\text{neutral}}= 
\sum_i\Omega_{\rm med}^{i}= 
-T \sum_i \sum_{s=\pm 1} \int \frac{d^3k}{(2\pi)^3} 
\biggl\{{\rm ln}
\left( 1+e^{-\beta (\epsilon^{i}_{k,s} - \mu^{*}_{i} )}\right) \\
+ {\rm ln}\left( 1+e^{-\beta (\epsilon^{i}_{k,s}+\mu^{*}_{i})}
\right) \biggr\},
\label{ther_pot_Fermi_sea_neutral_baryons}
\end{equation}
with the single-fermion energy for the $i^{th}$ neutral baryon given as
\begin{equation}
\epsilon^i_{k,s} = \sqrt{k_z^2 
+ \left ( \sqrt{k_x^2 +k_y^2+ {M_i^{*}}^2}-s\kappa_i B\right)^2}.
\end{equation}
For a given baryon density, $\rho_B=\sum_i \rho_i$, where $\rho_i$ denotes the number density of the $i^{\mathrm{th}}$ baryon, the isospin asymmetry parameter is defined as $\eta=-\frac{\sum_i I_{3i}\rho_i}{\rho_B}$, with $I_{3i}$ being the third component of the isospin quantum number of the $i^{\mathrm{th}}$ baryon. The minimization of the thermodynamic potential with respect to the scalar and vector fields yields a set of coupled nonlinear equations, which are solved self-consistently for different values of the magnetic field $B$, baryon density $\rho_B$, temperature $T$, and isospin asymmetry parameter $\eta$. We have,

\begin{eqnarray}
\frac{\partial \Omega}{\partial \sigma}=
\frac{\partial \Omega}{\partial \zeta}=
\frac{\partial \Omega}{\partial \delta} =
\frac{\partial \Omega}{\partial \chi} =
\frac{\partial \Omega}{\partial \omega} = 
\frac{\partial \Omega}{\partial \rho} = 0.
%\frac{\partial \Omega}{\partial \phi} & 
\end{eqnarray}
This results in the following system of coupled equation:
\begin{equation}
\begin{aligned}
k_0\chi^2\sigma
&-4k_1(\sigma^2+\zeta^2+\delta^2)\sigma
-2k_2(\sigma^3+3\sigma\delta^2)
-2k_3\chi\sigma\zeta  \\
&-\frac{\xi}{3}\chi^4\frac{2\sigma}{\sigma^2-\delta^2}
+\left(\frac{\chi}{\chi_0}\right)^2m_\pi^2f_\pi
-\left(\frac{\chi}{\chi_0}\right)^2
m_\omega\omega\frac{\partial m_\omega}{\partial\sigma} \\
&-\left(\frac{\chi}{\chi_0}\right)^2
m_\rho\rho\frac{\partial m_\rho}{\partial\sigma}
-\sum_{i=p,n}g_{\sigma i}\rho_s^i=0,
\end{aligned}
\label{sigmaeq}
\end{equation}
\begin{equation}
k_0\chi^2\zeta
-4k_1(\sigma^2+\zeta^2+\delta^2)\zeta
-4k_2\zeta^3
-k_3\chi(\sigma^2-\delta^2)
-\frac{\xi}{3}\chi^4\zeta
+\left(\frac{\chi}{\chi_0}\right)^2
\left(
\sqrt{2}m_K^2f_K
-\frac{1}{\sqrt{2}}m_\pi^2f_\pi
\right)=0,
\label{zetaeq}
\end{equation}
\begin{equation}
k_0\chi^2\delta
-4k_1(\sigma^2+\zeta^2+\delta^2)\delta
-2k_2(\delta^3+3\sigma^2\delta)
+2k_3\chi\delta\zeta
+\frac{2}{3}\xi\chi^4
\frac{\delta}{\sigma^2-\delta^2}
-\sum_{i=p,n}g_{\delta i}\rho_s^i=0,
\label{deltaeq}
\end{equation}
\begin{multline}
k_0\chi(\sigma^2+\zeta^2+\delta^2)
-k_3(\sigma^2-\delta^2)\zeta
+\chi^3\left(1+\ln\frac{\chi^4}{\chi_0^4}\right)
+(4k_4-\xi)\chi^3
-\frac{4}{3}\xi\chi^3
\ln\!\left[
\frac{(\sigma^2-\delta^2)\zeta}{\sigma_0^2\zeta_0}
\frac{\chi^3}{\chi_0^3}
\right] \\
-\frac{\chi}{\chi_0^2}
\left[
m_\pi^2f_\pi\sigma
+\left(
\sqrt{2}m_K^2f_K
-\frac{1}{\sqrt{2}}m_\pi^2f_\pi
\right)\zeta
+m_\omega^2\omega^2
+m_\rho^2\rho^2
\right]
=0,
\label{chieq}
\end{multline}

\begin{equation}
\frac{\chi^2}{\chi_0^2}m_\omega^2\omega
+4g_4\omega^3
+12g_4\omega\rho^2
-\sum_{i=p,n}g_{\omega i}\rho_i=0,
\label{omegaeq}
\end{equation}
\begin{equation}
\frac{\chi^2}{\chi_0^2}m_\rho^2\rho
+4g_4\rho^3
+12g_4\omega^2\rho
-\sum_{i=p,n}g_{\rho i}\rho_i=0,
\label{rhoeq}
\end{equation}
where $\rho_i$ represents the number density of $i^{\text{th}}$ baryon, $\rho_i=-\partial\Omega/\partial\mu_i^*$, and for charged baryon is expressed as
\begin{align}
\rho_i &= 
\frac{|q_i|B}{2\pi^2}
\bigg[
\sum_{\nu=0}^{\nu_{\max}}
\int_{0}^{\infty} \! dk_z \,
\left(
\frac{1}{1 + e^{\beta(\epsilon^{i}_{k,\nu,s=+1}-\mu_i^{*})}}
-
\frac{1}{1 + e^{\beta(\epsilon^{i}_{k,\nu,s=+1}+\mu_i^{*})}}
\right)
\notag\\[4pt]
&\quad
+
\sum_{\nu=1}^{\nu_{\max}}
\int_{0}^{\infty} \! dk_z \,
\left(
\frac{1}{1 + e^{\beta(\epsilon^{i}_{k,\nu,s=-1}-\mu_i^{*})}}
-
\frac{1}{1 + e^{\beta(\epsilon^{i}_{k,\nu,s=-1}+\mu_i^{*})}}
\right)
\bigg].
\end{align}
For neutral baryon it reduces to
\begin{align}
\rho_{i}=  \sum_{s = \pm 1}
 \int
\frac{d^3k}{(2\pi)^3}\Bigg[
\frac{1}{1+e^{\beta(  \epsilon^i_{k, s} 
-\mu^{*}_{i})}} 
-
\frac{1}{1+e^{\beta(  \epsilon^i_{k, s} 
+\mu^{*}_{i})}} 
\Bigg].
\label{rhovp_neutral}
\end{align}
In Eqs.~(\ref{sigmaeq})-(\ref{rhoeq}), the scalar density of the $i^\text{th}$ baryon, $\rho_s^i=\langle\bar\psi^i\psi^i\rangle=\partial\Omega/\partial m_i^*$ is calculated by neglecting the contribution from the Dirac sea to the thermodynamic potential. The expression of scalar density for $i^{\text{th}}$ charged baryon is given as 
\begin{align}
\rho_{s}^{i,\text{med}} &=
\frac{|q_i|B}{2\pi^2}\,M_i^{*}
\biggl[
\sum_{\nu=0}^{\infty}
\int_{0}^{\infty} dk_z\,
\frac{\sqrt{M_i^{*2}+2\nu|q_i|B} - \kappa_i B}
{\epsilon_{k,\nu,s=1}^{i}\sqrt{M_i^{*2}+2\nu|q_i|B}} \notag\\
&\quad
\left(
\frac{1}{1+e^{\beta(\epsilon_{k,\nu,s=1}^{i}-\mu_i^{*})}}
+
\frac{1}{1+e^{\beta(\epsilon_{k,\nu,s=1}^{i}+\mu_i^{*})}}
\right)+\sum_{\nu=1}^{\infty}
\int_{0}^{\infty} dk_z\,
\frac{\sqrt{M_i^{*2}+2\nu|q_i|B} + \kappa_i B}
{\epsilon_{k,\nu,s=-1}^{i}\sqrt{M_i^{*2}+2\nu|q_i|B}} \notag\\
&\quad
\left(
\frac{1}{1+e^{\beta(\epsilon_{k,\nu,s=-1}^{i}-\mu_i^{*})}}
+
\frac{1}{1+e^{\beta(\epsilon_{k,\nu,s=-1}^{i}+\mu_i^{*})}}
\right )
\Biggr].
\end{align}
For neutral baryons, the scalar density gets contributions from both the spin states. In the presence of an external magnetic field, the anomalous magnetic moment modifies the single-particle energies which leads to the following expression for the scalar density:
\begin{align}
\rho_{s}^{i,\rm med} &=
M_{i}^{*}
\sum_{s=\pm 1}
\int \frac{d^{3}k}{(2\pi)^{3}}
\frac{
\sqrt{k_{x}^{2} + k_{y}^{2} + M_{i}^{*2}} - s\kappa_{i}B
}{
\epsilon_{k,s}^{i}\sqrt{k_{x}^{2} + k_{y}^{2} + M_{i}^{*2}}
}
\Bigg[
\frac{1}{1+e^{\beta(\epsilon_{k,s}^{i}-\mu_{i}^{*})}}
+
\frac{1}{1+e^{\beta(\epsilon_{k,s}^{i}+\mu_{i}^{*})}}
\Bigg].
\label{eq:Fermi_neutral}
\end{align}
The magnetic field also modifies the vacuum structure of the system through baryonic Dirac sea polarization. Taking this effect into account, the renormalized Dirac sea contribution to the scalar density of the $i^{th}$ baryon can be expressed as\cite{PhysRevD.90.125036,PhysRevD.98.056024,Mishra:2023uhx}
 
\begin{align}
\rho_{s}^{\mathrm{DS},i}
=
-\frac{1}{4\pi^{2}}
\Biggl[
\frac{(q_i B)^{2}}{3 M_i^{*}}
+
(\kappa_i B)^{2} M_i^{*}
+
|q_i|B\,\kappa_i B
\Biggl\{
\frac{1}{2}
+
2\ln\!\left(\frac{M_i^{*}}{M_i}\right)
\Biggr\}
\Biggr].
\label{eq:DS}
\end{align}

\subsection{Light-Front Quark Model (LFQM)}
\label{subsec:LFQM}
\subsubsection{Light-Front Wave Function (LFWF) and Effective Hamiltonian}
The in-medium properties of the heavy-light mesons are calculated within the LFQM, which provides a relativistic description of mesons in terms of their constituent quark degrees of freedom. The meson properties are obtained from the QCD-inspired effective Hamiltonian as the well-known linear plus Coulomb interaction given by \cite{PhysRevD.59.074015,PhysRevD.80.054016,PhysRevD.110.014020}
\begin{align}
H_{q\bar{q}}=H_0 + V_{q\bar{q}}= \sqrt{m^2_q+{\vec
k}^2}+\sqrt{m^2_{\bar{q}}+{\vec k}^2}+V_{q\bar{q}},
\end{align}
where the potential $V_{q\bar{q}}$ is the interaction potential for the pseudoscalar and vector mesons and defined as
\begin{eqnarray}\label{eq:potential}
V_{q\bar{q}} &=& a + br -\frac{4\alpha_s}{3r} %\frac{b(1-{\rm e}^{-c r})}{c} \nonumber\\
+ \frac{32\pi \alpha_s \tilde{\delta}^3(r) }{9m_q m_{\bar{q}}} \langle{\bm{S}_q \cdot \bm{S}_{\bar{q}} }\rangle.
\end{eqnarray}
Here, the first three terms correspond to the constant, linear confinement, and Coulomb interactions, respectively, while the last term represents the hyperfine interaction essential to distinguish between the pseudoscalar ($0^{-+}$) and vector ($1^{--}$) mesons. The parameters a and b represent the constant energy shift and the strength of the linear confining potential (string tension), respectively, and $\alpha_s$ is the coupling constant, that measures the strength of the strong interaction. In the present work, $\alpha_s$ is treated as a constant. However, its modification in nuclear matter has been investigated in Ref.~\cite{PhysRevD.107.114010}, while its temperature dependence has been studied in Ref.~\cite{Schneider:2003uz}. The term $\langle{\bm{S}_q \cdot \bm{S}_{\bar{q}} }\rangle$ yields the values of $\frac{1}{4}$ and $-\frac{3}{4}$ for the vector and pseudoscalar mesons, respectively. Here, the spin–spin interaction is smeared using a Gaussian function as 
\begin{eqnarray}
    \tilde{\delta}^3(r) = \frac{\tilde{\Lambda}^3}{\pi^{3/2}} \mathrm{e}^{-\tilde{\Lambda}^2 r^2},
\end{eqnarray}
where $\tilde{\Lambda}$ determines the strength of smearing effect. 
\par To construct the meson bound state in the LFQM, the LFWF is expressed in terms of Lorentz-invariant internal variables, such as the longitudinal momentum fraction $x_i = p_i^+/P^+$, intrinsic transverse momentum ${\textbf{k}_{{\perp}{i}}}=\textbf{p}_{\perp i}-x_i\textbf{P}_{\perp}$, where $P^\mu = (P^+,P^-,\bm{P}_\perp)$ and $p^\mu_i$ denotes the four-momentum of the meson and the $i$-th constituent quark, respectively. Here we define the longitudinal momentum fraction $x \equiv x_q$ and the transverse momentum ${\bf k}_\perp \equiv {\bf k}_{\perp q}$.

We perform the variable transformation $(k_z, \textbf{k}_\perp) \to (x,\textbf{k}_\perp)$. 
In this case, $x$ can be related to $k_z$ as
%#######################################
\begin{eqnarray}\label{kz}
    x = \frac{ E_q - k_z}{E_q + E_{\bar{q}}}, \qquad  1 - x = \frac{E_{\bar{q}} + k_z}{E_q + E_{\bar{q}}},
\end{eqnarray}
%#######################################
where $E_q = \sqrt{m^2_q + \textbf{k}^2}$.
Therefore, longitudinal component $k_z^*$ is expressed as
%#######################################
\begin{eqnarray}\label{eq:k_z}
    k_z^* = \left( x - \frac{1}{2} \right) M_0^* + \frac{(m^{*2}_{\bar{q}} -m^{*2}_q)}{2M_0^*},
\end{eqnarray} 
%#################################
with the invariant meson mass $M_0 = E_q + E_{\bar{q}}$ is given as
%#######################################
\begin{eqnarray}
	M_0^{*2} = \frac{\textbf{k}_{\bot}^2 + {m^*_q}^2}{x}  + \frac{\textbf{k}_{\bot}^2 + {m^*_{\bar{q}}}^2}{1-x}.
\end{eqnarray} 
%#######################################
The complete LFWF can be factorized into a radial part and a spin-orbit part. While the radial wave function describes the momentum distribution of the constituent quarks, the spin-orbit wave function accounts for the spin structure of the meson. The LFWF of the ground-state meson can be written as \cite{PhysRevD.110.014020,PhysRevD.106.014009}
\begin{align}
\mathbf  \Psi_{100}^{JJ_z}(x_{},\mathbf{k}_{\perp})=\mathcal{R^*}_{\lambda_{q}\lambda_{\bar{q}}}^{JJ_z}(x_{i},\mathbf{k}_{\bot
  }) \mathbf \Phi^*(x_{},\mathbf{k}_{\bot }),
  \label{eqn:Wavefunction}
\end{align}
where $\mathbf{\Phi}^*(x,\textbf{k}_{\perp})$ represents the vacuum radial part to describe the internal quark-antiquark structure of a meson within the LFQM and $\mathcal{R^*}_{\lambda_{q}\lambda_{\bar{q}}}^{JJ_z}$ denotes the interaction-independent spin-orbit wave function. The covariant form  of $\mathcal{R^*}_{\lambda_{q}\lambda_{\bar{q}}}^{JJ_z}$ obtatained from the Melosh transformation \cite{PhysRevD.9.1095} is expressed as 
\begin{eqnarray}
	\mathcal{R^*}^{JJ_z}_{\lambda_q\lambda_{\bar{q}}} &=&  \frac{1}{\sqrt{2} \tilde{M}^*_0} 
	\bar{u}_{\lambda_q}^{}(p_q) \Gamma_{\rm M} v_{\lambda_{\bar{q}}}(p_{\bar{q}}),
\end{eqnarray}
with $\tilde{M}^*_0 \equiv \sqrt{{{M^*_0}}^2 - (m^*_q -m^*_{\bar{q}})^2}$ and $u(p_q)$ and $v(p_{\bar{q}})$ denotes the Dirac spinors for quark and antiquark respectively. The spin-orbit wave functions satisfy the relation $\Large\langle \mathcal R_{\lambda_q \lambda_{\bar{q}}}^{*\textit{J} \textit{J}_z}|\mathcal R_{\lambda_q \lambda_{\bar{q}}}^{*\textit{J} \textit{J}_z}\Large\rangle$ =1.
The radial wave function encodes the momentum distribution of the constituent quarks and plays a crucial role in determining meson properties. To apply the variational principle, we take the lowest-order harmonic oscillator wave function as the in-medium trial radial wave function for both pseudoscalar and vector mesons and is given by
%#######################################
\begin{eqnarray}
\mathbf \Phi^*_{1S}(x, \mathbf{k}_\perp) &=& \frac{4\pi^{3/4}} {\beta^{3/2}} \sqrt{\frac{\partial
{k_z}^*}{\partial x}} e^{-\frac{\mathbf{k}^2}{2\beta^2}}.
\end{eqnarray}
%#######################################
Here, $\beta$ is the variational parameter that characterizes the spatial extension of the meson in momentum space. The Jacobian factor $\frac{\partial k_z^*}{\partial x}$ arises from the transformation between the light-front variables $(x,\mathbf{k}_\perp)$ and the three-momentum variables $(\mathbf{k_\perp},k_z^*)$ which ensures the normalization of the wave function \cite{PhysRevD.80.054016}.
%#######################################

\par To obtain the meson mass, we employ the variational principle by minimizing the expectation value of the QCD-inspired effective Hamiltonian with respect to the variational parameter $\beta$. This procedure determines the optimal wave function corresponding to the ground-state meson within the chosen basis. Consequently, the mass eigenvalue is obtained by applying the minimizing condition
$M^*_{q\bar q}
=\langle \mathbf{\Psi}_{q\bar q}|H_{q\bar q}|\mathbf{\Psi}_{q\bar q}\rangle
=\langle \mathbf{\Phi}^{*}_{1S}|H_{q\bar q}|\mathbf{\Phi}^{*}_{1S}\rangle,
$ where the spin-orbit wave function is normalized independently, reducing the expectation value to the radial part. The masses obtained from the previous variational procedure represent the vacuum meson masses ($eB=0$) prior to the inclusion of magnetic-field effects. The presence of an external magnetic field leads the charged $D$ and $D^*$ mesons masses to have additional contribution from Landau levels. For a charged pseudoscalar and vector $D$ mesons, the masses in a magnetic field is given as \cite{Gubler:2015qok}
\begin{eqnarray}
    M_D^{\rm eff}(n,p_z)=\sqrt{{M_P^*}^2+(2n+1)eB+p_z^2},
\label{eq:PS_LLL}    
\end{eqnarray}
and
\begin{eqnarray}
    M_{D^*}^{\rm eff}(n,p_z,S_z)=\sqrt{{M_V^*}^2+(2n+1)eB+p_z^2+gS_zeB},
\label{eq:vectorLLL}    
\end{eqnarray}
respectively. Here, $M_P^*$ and $M_V^*$ denote the in-medium masses of  pseudoscalar and vector mesons, respectively, obtained from the variational solution of Hamiltonian described above. Furthermore, $n$ and $p_z$ denotes the Landau level index and the momentum along the magnetic field direction, respectively, while $g$ is gyromagnetic ratio and its value is taken to be $2$. For simplicity, we take the case of zero momentum in $z$-direction ($p_z=0$). Since the higher Landau levels only contribute to the sufficiently large magnetic fields, we restrict our present analysis only to lowest Landau level (LLL) ($n=0$) case. In present work, we have incorporated the effects of Landau quantization through the above dispersion relations, and the resulting masses are subsequently employed in PV mixing formalism.

\subsubsection{Effect of PV mixing }
The presence of an external magnetic field reduces the rotational symmetry of the system from $SO(3)$ to the subgroup corresponding to the rotations about the magnetic field direction. This results in the coupling of the pseudoscalar ($J^P=0^-$) meson to its longitudinal ($S_z=0$) polarization of its corresponding vector ($J^P=1^-$) counterpart \cite{PhysRevLett.113.172301, PhysRevD.91.045025,Alford:2013jva, Bonati:2015dka}. This magnetic field induced reduction modifies the physical masses of both mesons through level repulsion. \par
The interaction responsible for the mixing is described through the effective Lagrangian \cite{PhysRevD.91.045025}
\begin{eqnarray}
    \mathcal{L}_{PV}= \frac{g_{PV}}{M_{\rm av}}e\tilde{F}_{\mu \nu}(\partial^\mu P)V^\nu,
\end{eqnarray}
where $P$ and $V^\mu$ denote the pseudoscalar and vector meson field, respectively, $M_{\rm av}=(M_D^{\rm eff}+M_{D^*}^{\rm eff})/2$ is their average mass, 
$\tilde{F}_{\mu \nu}$ is the dual electromagnetic field tensor, and $g_{PV}$ represents the effective coupling governing the radiative transition between pseudoscalar and vector states. Diagonalization of the resulting mass matrix yields the physical masses in the magnetic field. The physical mass eigenvalue is given as \cite{PhysRevD.91.045025}
\begin{equation}
M_{D^{*},D}^{2}
=
\frac{1}{2}
\left(
M_{+}^{2}
+
\frac{\gamma^{2}}{M_{\rm av}^{2}}
\pm
\sqrt{
M_{-}^{4}
+
\frac{2\gamma^{2}M_{+}^{2}}{M_{\rm av}^{2}}
+
\frac{\gamma^{4}}{M_{\rm av}^{4}}
}
\right),
\label{eq:PVmass}
\end{equation}
with $M_\pm^2={M_{D^*}^{\rm eff}}^2\pm {M_D^{\rm eff}}^2$ and $\gamma=g_{PV}eB$. Thus, the combined effects of Landau quantization and PV mixing are incorporated into the meson masses, which are subsequently used to calculate the medium-dependent masses.

\subsubsection{Weak Decay Constant and Distribution Amplitudes (DAs)}
The decay constants of pseudoscalar and vector mesons are fundamental quantities that determine the strength of the meson coupling to the corresponding quark currents. They play an important role in the description of weak decay processes and provide valuable information about the internal structure of hadrons. For a pseudoscalar meson with four-momentum $P^\mu$ and a vector meson with polarization $(J_z)$, the decay constants $f_P$ and $f_V$ are defined through the matrix elements of the axial-vector and vector currents, respectively \cite{PhysRevD.107.114010,PhysRevD.75.073016},
\begin{eqnarray}
\langle 0 |\bar q \gamma^\mu \gamma_5 q | P(\textit{P}) \rangle & = & i f_P \textit{P}^\mu ,\\
\langle 0 |\bar q \gamma^\mu q | V(\textit{P},  \textit{J}_z) \rangle & = & f_V \textit{M}_V \epsilon^\mu (\textit{J}_z),
\end{eqnarray}
where $M_V$ and $\epsilon^\mu(J_z)$ represent the mass and polarization vector of the vector meson, respectively. Within the LFQM, the medium dependence of the decay constants arises through modifications of the constituent quark masses and the corresponding LFWF. By integrating over the internal momentum variables, the in-medium decay constants can be expressed as 
\begin{eqnarray}
\label{eq:PS_DC}
f^*_P & = & 2\sqrt{6} \int_0^1 dx \int \frac{d^2 \mathbf{k}_\perp}{2(2\pi)^3} \frac{\mathbf{\Phi}^*(x, \mathbf{k}_\perp) }{\sqrt{\mathcal{A}^{*2} + \mathbf{k}_\perp^2} }\ \mathcal{A^*},\\
f^*_V & = & 2\sqrt{6} \int_0^1 dx \int \frac{d^2 \mathbf{k}_\perp}{2(2\pi)^3} \frac{\mathbf{\Phi}^*(x, \mathbf{k}_\perp) }{\sqrt{\mathcal{A}^{*2} + \mathbf{k}_\perp^2} }\left(\ \mathcal{A^*}+\frac{2\textbf{k}^2_\perp}{\mathcal{M}^*}\right).
\label{eq:Vector_DC}
\end{eqnarray}
Another important quantity that provides insight into the mesonic structure is the distribution amplitude (DA). The leading-twist DA describes the longitudinal momentum distribution of the valence quarks inside a meson and serves as an essential nonperturbative input in the study of hard exclusive processes. Physically, the twist-2 DA corresponds to the probability amplitude of finding a hadron in its lowest Fock-state configuration with a small transverse separation between its constituents \cite{Chernyak:1983ej,Lepage:1980fj}. The leading-twist DAs for pseudoscalar and vector mesons are respectively defined through the following nonlocal light-cone matrix elements \cite{PhysRevD.107.114010,PhysRevD.75.034019,PhysRevD.96.016022}
\begin{eqnarray}
A_{P}^+ &=& \langle 0 | \bar{q}(z) \gamma^+ \gamma_5 q(-z) | P(\textit{P}) \rangle \nonumber \\
&=& i f_P \textit{P}^+ \int_0^1 dx \, e^{i (2x-1) \mathbf{\textit{P}} \cdot z} \, \phi_P(x)\Big|_{z^+ = z_\perp = 0}, \\
A_{V}^+ &=& \langle 0 | \bar{q}(z) \gamma^+ q(-z) | V(\textit{P}, 0) \rangle \nonumber \\
&=& f_V \textit{M}_V \epsilon^+(0) \int_0^1 dx \, e^{i (2x-1) \textit{P} \cdot z} \, \phi_V(x)\Big|_{z^+ = z_\perp = 0}.
\end{eqnarray}
In the LFQM, the DAs are directly related to the corresponding LFWFs. The medium modifications enter through the in-medium wave function and effective constituent quark masses. By integrating out the transverse momentum dependence of the LFWF, one obtains the longitudinal momentum distribution of the constituent quarks. Consequently, the medium-modified DAs for pseudoscalar, $\phi^*_P(x)$, and vector mesons, $\phi^*_V(x)$, can be written as
\begin{eqnarray}
\label{eq:7}
\phi^*_P(x) &=& \frac{2\sqrt{6}}{ f^*_P} \int \frac{d^2 \mathbf{k}_\perp}{2(2\pi)^3} \frac{\mathbf \Phi^*(x, \mathbf{k}_\perp)}{\sqrt{\mathcal A^{*2} + \mathbf{k}_\perp^2}}\mathcal{A^*},\\
\phi^*_V(x) &=& \frac{2\sqrt{6}}{ f^*_V} \int \frac{d^2 \mathbf{k}_\perp}{2(2\pi)^3} \frac{\mathbf \Phi^*(x, \mathbf{k}_\perp)}{\sqrt{\mathcal A^{*2} + \mathbf{k}_\perp^2}}\left(\mathcal{A^*}+\frac{2 \mathbf{k^2_\perp}}{\mathcal{M}^*}\right).
\label{eq:8}
\end{eqnarray}
The above expressions establish a direct connection between the distribution amplitudes and the underlying LFWF. They provide a useful framework for investigating how the surrounding nuclear medium influences the momentum distribution and internal structure of pseudoscalar and vector mesons.

 \section{Results and Discussion}
 \label{sec:results}

\renewcommand{\thetable}{\arabic{table}}
%table
\begin{table}[htbp]
\centering
\renewcommand{\arraystretch}{1.25}
\setlength{\tabcolsep}{6pt}
\resizebox{\textwidth}{!}{%
\begin{tabular}{|c|c|c|c|c|c|c|c|c|}
\noalign{\hrule height 1.5pt}
$k_{0}$ & $k_{1}$ & $k_{2}$ & $k_{3}$ & $k_{4}$ & $g_{s}$ & $g_{v}$ & $g_{4}$ & $\rho_{0}$  (fm$^{-3}$)\\
\hline
4.94 & 2.12 & $-10.16$ & $-5.38$ & $-0.06$ & 3.85 & 9.14 & 37.5 & 0.16 \\
\hline
$\sigma_{0}$ (MeV) & $\zeta_{0}$ (MeV) & $\chi_{0}$ (MeV)  & $f_{\pi}$ (MeV) & $m_\pi$ (MeV) & $m_{K}$ (MeV) & $f_{K}$ (MeV) & $m_{\omega}$ (MeV) & $m_{\phi}$ (MeV) \\
\hline
$-93$ & $-96.87$ & 254.38 & 93 & 139 & 496 & 115 & 783 & 1020 \\
\hline
$g_{\sigma}^{u}$ & $g_{\sigma}^{d}$ & $g_{\sigma}^{s}$ & $g_{\zeta}^{u}$ & $g_{\zeta}^{d}$ & $g_{\zeta}^{s}$ & $g_{\delta}^{u}$ & $g_{\delta}^{d}$ & $g_{\delta}^{s}$ \\
\hline
2.72 & 2.72 & 0 & 0 & 0 & 3.85 & 2.72 & 2.72 & 0 \\
\hline
$g_{\omega}^{u}$ & $g_{\omega}^{d}$ & $g_{\omega}^{s}$ & $g_{\phi}^{u}$ & $g_{\phi}^{d}$ & $g_{\phi}^{s}$ & $g_{\rho}^{u}$ & $g_{\rho}^{d}$ & $g_{\rho}^{s}$ \\
\hline

% \hline
% $g_{\omega}^{u}$ & $g_{\omega}^{d}$ & $g_{\omega}^{s}$ & $g_{\phi}^{u}$ & $g_{\phi}^{d}$ & $g_{\phi}^{s}$ & $g_{\rho}^{u}$ & $g_{\rho}^{d}$ & $g_{\rho}^{s}$ \\
% \hline

3.23 & 3.23 & 0 & 0 & 0 & 4.57 & 3.23 & 3.23 & 0 \\

\hline
$g_{\sigma_p}$ & $g_{\zeta p}$ & $g_{\delta p}$ & $g_{\sigma_n}$ & $g_{\zeta n}$ & $g_{\delta n}$ & $g_{\omega_p}$ & $g_{\rho p}$ & $g_{\phi _p}$ \\
\hline
6.64 & 0 & 2.72 & 6.64 & 0 & 2.72 & 9.69 & 8.89 & 0 \\
\hline

$g_{\omega_n}$ & $g_{\rho n}$ & $g_{\phi n}$ & $\xi$ & -- & -- & -- & -- & -- \\
\hline
9.69 & 8.89 & 0 & $0.182$ & -- & -- & -- & -- & -- \\

\noalign{\hrule height 1.5pt}

\end{tabular}
}%
\caption{List of model parameters used in the present work.}
\label{tab:1}
\end{table}
In this section, we present the results of in-medium properties of pseudoscalar and vector $D$ mesons in the presence of an external magnetic field and different baryon densities. We examine the in-medium masses, weak decay constants, and DAs of the pseudoscalar ($D^0, D^+, D_s$) and vector ($D^{0*}, D^{+*}, D_s^*$) mesons using the LFQM modulated by medium-modified quark masses calculated using CQMF. The model parameters, which are used to solve the equations of motion are shown in Table~\ref{tab:1}. The constituent quark masses in free space (in GeV) and the corresponding potential parameters were obtained by fitting the ground-state mass spectra of the $D^0$ and $D^{0*}$ mesons. The extracted values are listed in Table~\ref{tab:parameters} and the $\beta$ parameters for the pseudoscalar \(D\) and vector \(D^*\) mesons computed by the variational principle (in the units of GeV) are given in Table~\ref{tab:2}.

\begin{table}[htbp]
\centering
\renewcommand{\arraystretch}{1.3}
\setlength{\tabcolsep}{20pt}

\begin{tabular}{|c|c|c|}
\hline
\textbf{Parameter} & \textbf{Value} & \textbf{Unit} \\
\hline
$m_u=m_d$ & 0.256 & GeV \\
\hline
$m_s$ & 0.457 & GeV \\
\hline
$m_c$ & 1.27 & GeV \\
\hline
$a$ & -0.043 & GeV \\
\hline
$b$ & 0.18 & GeV$^2$ \\
\hline
$\alpha_s$ & 0.565 & -- \\
\hline
\end{tabular}
\caption{Constituent quark masses and fitted potential parameters used in the present calculations \cite{10.1093/ptep/ptaf135}.}
\label{tab:parameters}
\end{table}

\begin{table}[htbp]
\centering
\renewcommand{\arraystretch}{1.5}
\setlength{\tabcolsep}{18pt}   % try 16pt, 18pt or 20pt

\begin{tabular}{|c|c|c|c|}
\hline
$J^{PC}$ & $\beta_{c\bar{u}}$ & $\beta_{c\bar{d}}$ & $\beta_{c\bar{s}}$ \\
\hline
$0^{-+}$ & 0.5331 & 0.5189 & 0.6066 \\
\hline
$1^{--}$ & 0.4896 & 0.4681 & 0.4742 \\
\hline
\end{tabular}

\caption{Gaussian parameter $\beta$ for pseudoscalar mesons ($D^0$, $D^+$, $D_s$) and vector mesons ($D^{0*}$, $D^{+*}$, $D_s^*$), determined using the variational principle~\cite{10.1093/ptep/ptaf135}.}
\label{tab:2}
\end{table}
%================ MASS =================%
\subsection{In-medium Masses of $D$ and $D^*$ Mesons}

In this subsection, we present the in-medium masses of the pseudoscalar and vector $D$ mesons. The medium-modified constituent quark masses are first calculated within the CQMF model and then used in the LFQM to obtain the in-medium $D$ mesons masses in magnetized isospin asymmetric nuclear matter. While the effect of magnetic field induced PV mixing is taken into account for both neutral and charged mesons, Landau quantization contribution is incorporated only for the charged states. The PV mixing is implemented through the effective coupling constant $g_{PV}=3.6736$ for the $D^0-D^{0*}$ system and $g_{PV}=0.9366$ for the $D^+-D^{+*}$ system \cite{Gubler:2015qok}. The value of the $g_{PV}$ for $D_s-D_s^*$ system is calculated using the same formalism as stated in Ref.~\cite{Gubler:2015qok} and the required value of the radiative decay width for $D_s^*$ meson $\Gamma[D_s^* \ \rightarrow D_s+\gamma]$ is 0.066 keV \cite{Donald:2013sra}. The calculated ground-state masses of the pseudoscalar ($D^0$, $D^+$, $D_s$) and vector ($D^{0*}$,$D^{+*}$, $D_s^*$) mesons are presented in Table~\ref{tab:3}. We compare our results with the experimental values reported by the Particle Data Group (PDG) \cite{PhysRevD.110.030001} as well as with previous theoretical studies based on the LFQM \cite{PhysRevC.92.055203} and the QCD sum rules \cite{QCD}. The masses of the non-strange pseudoscalar mesons are in good agreement with the experimental values, while moderate deviations are observed for the strange and vector mesons, with differences ranging from approximately 19 to 43 MeV. 
\begin{table*}[t]
    \centering
    \renewcommand{\arraystretch}{1.5}
    \begin{tabular}{| l | c c c c c c|}
        \hline
        \multirow{2}{*}{} & \multicolumn{6}{c|}{\textbf{Mass of Meson (GeV)}} \\
        \cline{2-7}
        & $M_{D^0}$ & $M_{D^+}$ & $M_{D_s}$ & $M_{D^{0*}}$ & $M_{D^{+*}}$ & $M_{D_s^*}$ \\
        \hline
        Present Work & 1.865 & 1.868 & 2.011 & 1.988 & 2.0099 & 2.087 \\
        \hline
        PDG\cite{PhysRevD.110.030001} & 1.865 & 1.869 & 1.968 & 2.007 & 2.010 & 2.112 \\
        \hline
        LFQM\ \cite{PhysRevC.92.055203} & 1.875 & {-} & 1.981 & 1.962 & {-} & 2.031 \\
        \hline
        QCD sum rules\ \cite{QCD} & $1.87 \pm 0.10$ & {-} & $1.97 \pm 0.10$ & $2.01 \pm 0.08$ & {-} & $2.11 \pm 0.07$ \\
        \hline
    \end{tabular}
    \caption{Predicted ground-state mass spectra (in GeV) of pseudoscalar ($D^0$, $D^+$, $D_s$) and vector ($D^{0*}$, $D^{+*}$, $D_s^*$) mesons, compared with experimental PDG data  and other theoretical model predictions.}
    \label{tab:3}
\end{table*}

\begin{figure*}[htbp]
    \centering
    \includegraphics[width=0.90\linewidth]{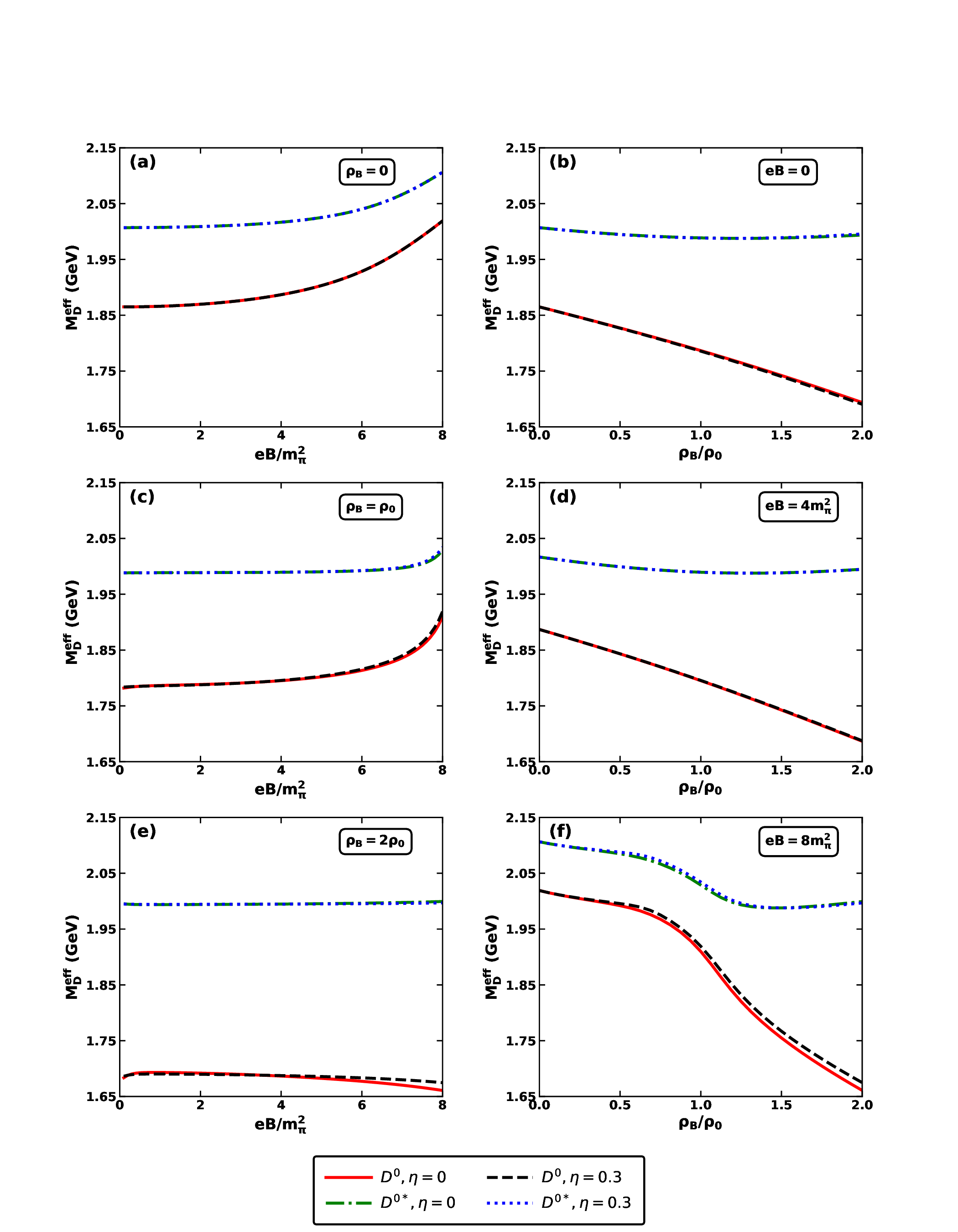}
    \caption{Variation of the in-medium effective masses of neutral $D$ mesons ($D^0$ and $D^{0*}$) as a function of $eB/m_\pi^2$ at fixed baryon density $\rho_B=0,\rho_0,2\rho_0$ [in subplots (a),(c),(e)] and as a function of $\rho_B/\rho_0$ at fixed magnetic field strength $eB/m_\pi^2=0,4m_\pi^2,8m_\pi^2$ [in subplots (b),(d),(f)] for isospin asymmetry $\eta = 0$ and $\eta = 0.3$ at $T=0.1$ GeV.}
    \label{fig:mass_D0}
\end{figure*}

The in-medium masses of $D$ and $D^*$ mesons are calculated in asymmetric nuclear matter in the presence of external magnetic field. These modifications in the masses of $D$ and $D^*$ meson in the nuclear matter arise due to the interactions of the light quarks present in the open charm meson with the scalar meson ($\sigma, \zeta$ and $\delta$). The variation of the medium-modified masses of neutral $D^0$ and $D^{0*}$ mesons for different values of isospin asymmetry $\eta=0,0.3$ is shown in Fig.~\ref{fig:mass_D0}. The subplots (a), (c), and (e) show the variation of masses with the magnetic field strength $eB/m_\pi^2$, whereas subplots (b), (d), and (f) give the variation with the baryonic density $\rho_B/\rho_0$. At $\rho_B=\rho_0$ and $T = 0.1$ GeV, the masses of both $D$ mesons increase with increasing isospin asymmetry. For $eB=8m_\pi^2$, the masses of the $D^0$ ($D^{0*}$) mesons increase from $1.9098\ (2.0288)$ GeV at $\eta=0$ to $1.9190\ (2.0342)$ GeV at $\eta=0.3$. As the baryon density is increased to $2\rho_0$, the corresponding masses decrease to $1.6607\ (1.9994)$ GeV in symmetric matter and change to $1.6747\ (1.9968)$ GeV for $\eta=0.3$. At zero baryon density, the masses of both mesons increase monotonically with increasing magnetic field, however the enhancement is more pronounced for the pseudoscalar $D^0$ meson than its vector counterparts. As the baryonic density increases from $\rho_0$ to $2\rho_0$, the overall masses decrease, while the magnetic field dependence becomes progressively weaker. This demonstrates that density-induced medium modification dominates over the magnetic field effect. In contrast, the subplots (b), (d), and (f) show that the masses decrease with the increasing baryonic density for all magnetic field strengths considered. We see that this reduction is significantly larger for pseudoscalar meson than its vector states over the entire density range. Furthermore, an increase in the magnetic field shifts the meson masses to higher values at a fixed density, whereas the effect of isospin asymmetry remains relatively small throughout our investigated space. It is also seen that the vector meson mass is larger than that of the pseudoscalar meson because of the spin-dependent hyperfine interaction between the quark and antiquark, which lowers the energy of the spin-singlet $(S=0)$ state relative to the spin-triplet $(S=1)$ state.  \par

Since the $D^0$ and $D^{0*}$ mesons carry no net electric charge, they cannot occupy discrete Landau orbits. Consequently, all the changes arise solely from the magnetic field effect. The external magnetic field polarizes the Dirac sea of the light quarks and strengthens the chiral condensates, known as magnetic catalysis, leading to an increase in the constituent light-quark masses. This consequently raises the masses of both the pseudoscalar and vector mesons within the LFQM. This growth is modest at low magnetic field but accelerates visibly beyond  $eB\sim5m_\pi^2$: this is because the Dirac sea correction to the scalar density in Eq.~(\ref{eq:DS}) carries a term proportional to $(eB)^2$, which becomes significant at strong fields. Thus, the neutral $D$ meson masses exhibit a nonlinear dependence on the magnetic field at sufficiently large $eB$ values. On the other hand, increasing baryonic density has the opposite effect: it drives the in-medium mass of $\bar{u}$ quark ($m_u^*$) back down as the scalar condensates weaken due to partial restoration of chiral symmetry, resulting in a substantial decrease in meson mass. This also explains the comparatively weaker magnetic field dependence observed at $\rho_B=2\rho_0$ (subplot (e)). The effective quark mass $m_q^*$ in Eq.~(\ref{eq:quark_mass}) is determined by the total scalar density, which receives contribution from both the field-dependent Dirac sea [Eq.~(\ref{eq:DS})] and the Fermi sea [Eq.~(\ref{eq:Fermi_neutral})]. At higher baryon density, $\rho_B=2\rho_0$, the Fermi sea contribution dominates and substantially suppresses the scalar condensates, while the Dirac sea contribution provides only a comparatively small correction. Consequently, the magnetic field dependence of the quark masses is considerably weakened, resulting in a reduced sensitivity of the $D$ and $D^*$ meson masses to the external magnetic field. We also observe that the decrease in effective meson masses with increasing baryon density is linear at low magnetic field, $eB=0$ and $4m_\pi^2$ [subplots~(b) and (d)], whereas it becomes clearly nonlinear at $eB=8m_\pi^2$ [subplot~(f)]. At lower magnetic fields, the Dirac sea contribution to the total scalar density is relatively small, so the density dependence is governed mainly by the Fermi sea term in Eq.~(\ref{eq:Fermi_neutral}), which decreases smoothly with $\rho_B$. In contrast, at $eB=8m_\pi^2$, the Dirac sea contribution enhances and actively competes with the density-dependent Fermi sea contribution. As the baryon density increases, the Fermi sea contribution gradually dominates, resulting in a nonuniform variation of effective quark masses and consequently a nonlinear decrease in the effective meson masses. The comparatively small effect of isospin asymmetry indicates that the contribution from the scalar-isovector $\delta$ field remains much weaker than that of the dominant $\sigma$ field. Consequently, moving from symmetric ($\eta=0$) to neutron-rich ($\eta=0.3$) matter pushes the mass of $D^0$ mesons upward relative to the symmetric-matter case. This effect is small at $\rho_B=0$ but grows steadily with density, becoming more pronounced at $\rho_B=2\rho_0$. Within the LFQM, the pseudoscalar and vector mesons differ only through the spin-dependent hyperfine interaction. Since the in-medium modification enters primarily through the constituent light-quark mass, the hyperfine contribution changes only weakly with density and magnetic field. Consequently, the vector meson exhibits a much smaller variation than the pseudoscalar state. 

\begin{figure*}[htbp]
    \centering
    \includegraphics[width=0.85\linewidth]{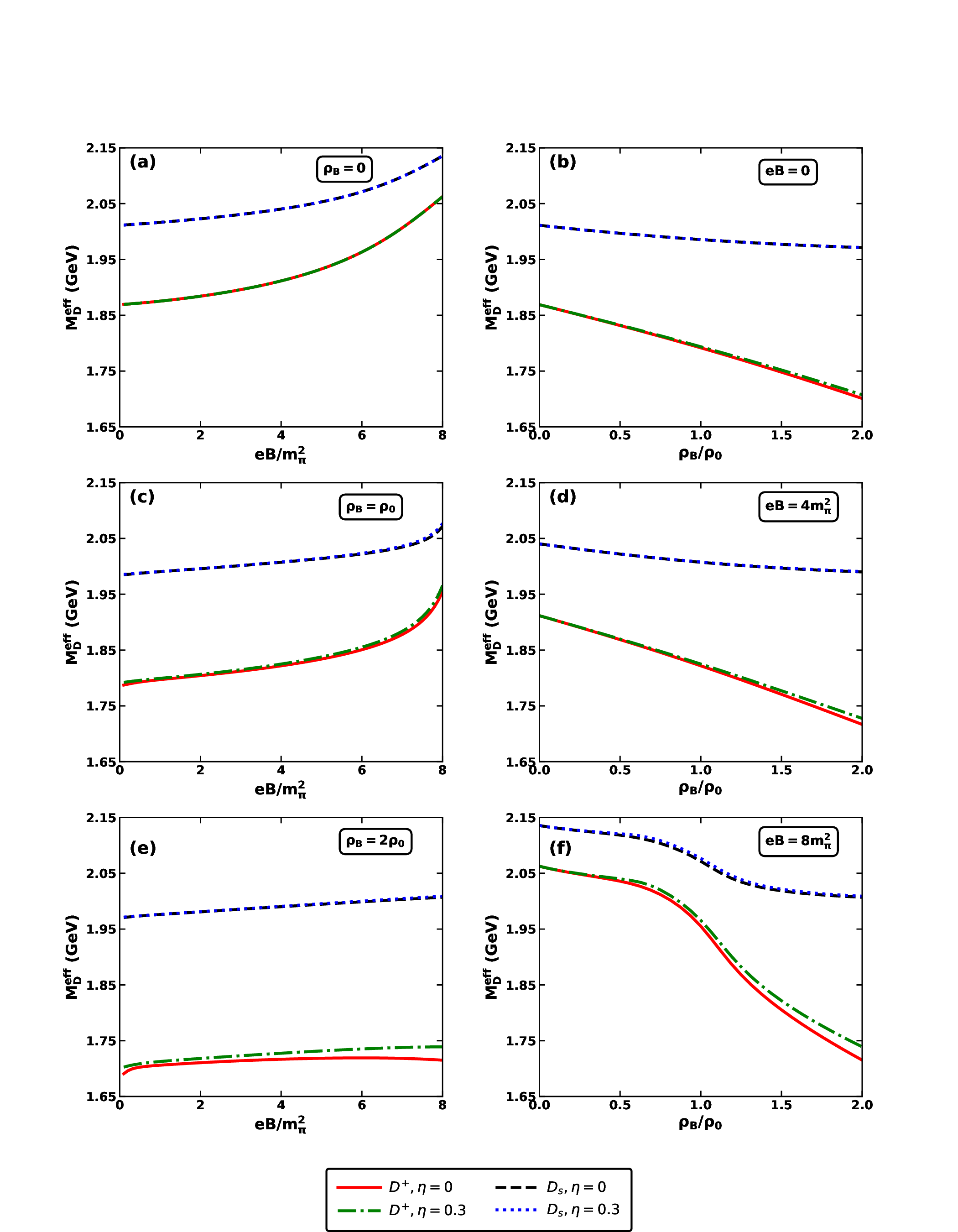}
    \caption{Variation of the in-medium effective masses of pseudoscalar charged $D$ mesons ($D^+$, $D_s$) as a function of $eB/m_\pi^2$ at fixed baryon density $\rho_B=0,\rho_0,2\rho_0$ [in subplots (a),(c),(e)] and as a function of $\rho_B/\rho_0$ at fixed magnetic field strengths $eB=0,4m_\pi^2,8m_\pi^2$ [in subplots (b),(d),(f)] for isospin asymmetry $\eta = 0$ and $\eta = 0.3$ at $T=0.1$ GeV. The effect of LLL quantization is considered.}
    \label{fig:mass_Ds}
\end{figure*}

\begin{table*}[t]
\centering
\renewcommand{\arraystretch}{1.25}
\setlength{\tabcolsep}{7pt}

\begin{tabular}{|c|c|c|c|c|c|c|}
\hline
\multirow{2}{*}{\textbf{Meson}} &
\multicolumn{2}{c|}{$\rho_B=0$} &
\multicolumn{2}{c|}{$\rho_B=\rho_0$} &
\multicolumn{2}{c|}{$\rho_B=2\rho_0$} \\
\cline{2-7}

& $eB=0$ & $eB=8m_\pi^2$
& $eB=0$ & $eB=8m_\pi^2$
& $eB=0$ & $eB=8m_\pi^2$ \\
\hline

% \multicolumn{7}{|c|}{\textbf{Pseudoscalar mesons}}\\
% \hline

$D^{+}(S_z=0)$ & 1.8689 & 2.0625 & 1.7915 & 1.9551 & 1.7008 & 1.7150 \\

\hline

$D_{s}(S_z=0)$ & 2.0111 & 2.1354 & 1.9855 & 2.071 & 1.9711 & 2.0070 \\

\hline

% \multicolumn{7}{|c|}{\textbf{Vector mesons}}\\
% \hline

$D^{+*}(S_z=-1)$ & 2.0099 & 2.077 & 1.9896 & 1.9951 & 1.9926 & 1.9583 \\
$D^{+*}(S_z=0)$  & 2.0099 & 2.1507 & 1.9896 & 2.0717 & 1.9926 & 2.0363 \\
$D^{+*}(S_z=+1)$ & 2.0099 & 2.222 & 1.9896 & 2.1456 & 1.9926 & 2.1114 \\

\hline

$D_s^{*}(S_z=-1)$ & 2.0870 & 2.1556 & 2.0535 & 2.0948 & 2.0339 & 2.0389 \\
$D_s^{*}(S_z=0)$  & 2.0870 & 2.2268 & 2.0535 & 2.1679 & 2.0339 & 2.1139 \\
$D_s^{*}(S_z=+1)$ & 2.0870 & 2.2957 & 2.0535 & 2.2387 & 2.0339 & 2.1864 \\

\hline

\end{tabular}

\caption{Calculated in-medium masses (GeV) of the charged pseudoscalar $D$ and vector $D^*$ mesons in symmetric nuclear matter ($\eta=0$) at $T=0.1$ GeV. The charged meson masses include the contribution from the lowest Landau level.}

\label{tab:eta0}

\end{table*}

\begin{table*}[t]
\centering
\renewcommand{\arraystretch}{1.50}
\setlength{\tabcolsep}{10pt}

\begin{tabular}{|c|c|c|c|c|c|c|}
\hline
\multirow{2}{*}{\textbf{Meson}} &
\multicolumn{2}{c|}{$\rho_B=0$} &
\multicolumn{2}{c|}{$\rho_B=\rho_0$} &
\multicolumn{2}{c|}{$\rho_B=2\rho_0$} \\
\cline{2-7}

& $eB=0$ & $eB=8m_\pi^2$
& $eB=0$ & $eB=8m_\pi^2$
& $eB=0$ & $eB=8m_\pi^2$ \\
\hline

% \multicolumn{7}{|c|}{\textbf{Pseudoscalar mesons}}\\
% \hline

$D^{+}(S_z=0)$ & 1.8689 & 2.0625 & 1.7935 & 1.9655 & 1.7075 & 1.7389 \\

\hline

$D_{s}(S_z=0)$ & 2.0111 & 2.1354 & 1.9857 & 2.0767 & 1.9714 & 2.0089 \\

\hline

% \multicolumn{7}{|c|}{\textbf{Vector mesons}}\\
% \hline

$D^{+*}(S_z=-1)$ & 2.0099 & 2.077 & 1.9901 & 2.0018 & 1.9906 & 1.9542 \\
$D^{+*}(S_z=0)$  & 2.0099 & 2.1507 & 1.9901 & 2.0782 & 1.9906 & 2.0324 \\
$D^{+*}(S_z=+1)$ & 2.0099 & 2.222 & 1.9901 & 2.1518 & 1.9906 & 2.1077 \\

\hline

$D_s^{*}(S_z=-1)$ & 2.0870 & 2.1556 & 2.0550 & 2.0996 & 2.0365 & 2.0404 \\
$D_s^{*}(S_z=0)$  & 2.0870 & 2.2268 & 2.0550 & 2.1726 & 2.0365 & 2.1154 \\
$D_s^{*}(S_z=+1)$ & 2.0870 & 2.2957 & 2.0550 & 2.2432 & 2.0365 & 2.1878 \\

\hline

\end{tabular}

\caption{Calculated in-medium masses (GeV) of the charged pseudoscalar $D$ and vector $D^*$ mesons in isospin asymmetric ($\eta=0.3$) nuclear matter at $T=0.1$ GeV. The charged meson masses include the contribution from the lowest Landau level.}

\label{tab:eta0.3}

\end{table*}

The corresponding results for the charged pseudoscalar $D^+ $ and $D_s$ mesons, including the LLL contribution are presented in Fig.~\ref{fig:mass_Ds}. The representative numerical values are listed in Tables~\ref{tab:eta0} and \ref{tab:eta0.3} for both $\eta=0$ and $0.3$. Similar to the neutral mesons discussed in Fig.~\ref{fig:mass_D0}, the effective masses increase with increasing magnetic field and decrease with the increasing baryonic density. However, unlike the neutral mesons, the charged states undergo Landau quantization, and their effective masses are obtained by including the LLL contribution through the dispersion relation given in Eq.~(\ref{eq:PS_LLL}). This results in a more pronounced increase in effective masses with increasing magnetic field than that of the neutral mesons. This distinct behavior of the charged mesons originates from the Landau quantization which raises the energy of the charged states and modifies the medium through changes in condensates and scalar field that leads to increase in effective masses. We further see that the $D_s$ meson undergoes comparatively weaker medium modifications than the $D^+$ meson over the entire density and magnetic field range. This behavior originates from the different scalar fields governing the constituent quark masses in the CQMF model. The light quark in $D^+$ meson couples to the non-strange scalar field $\sigma$, which decreases noticeably with increasing baryonic density, while strange $s$ quark in $D_s$ meson couples to the strange scalar field $\zeta$, whose medium modification is comparatively weaker. Consequently, the strange $s$ quark masses decreases slightly, resulting in a comparatively smaller shift in the $D_s$ effective mass. The effect of isospin asymmetry remains relatively weak because the contribution from the scalar-isovector $\delta$ field is much smaller than that of the dominant scalar-isoscalar fields.

\begin{figure*}[htbp]
    \centering
    \includegraphics[width=0.85\linewidth]{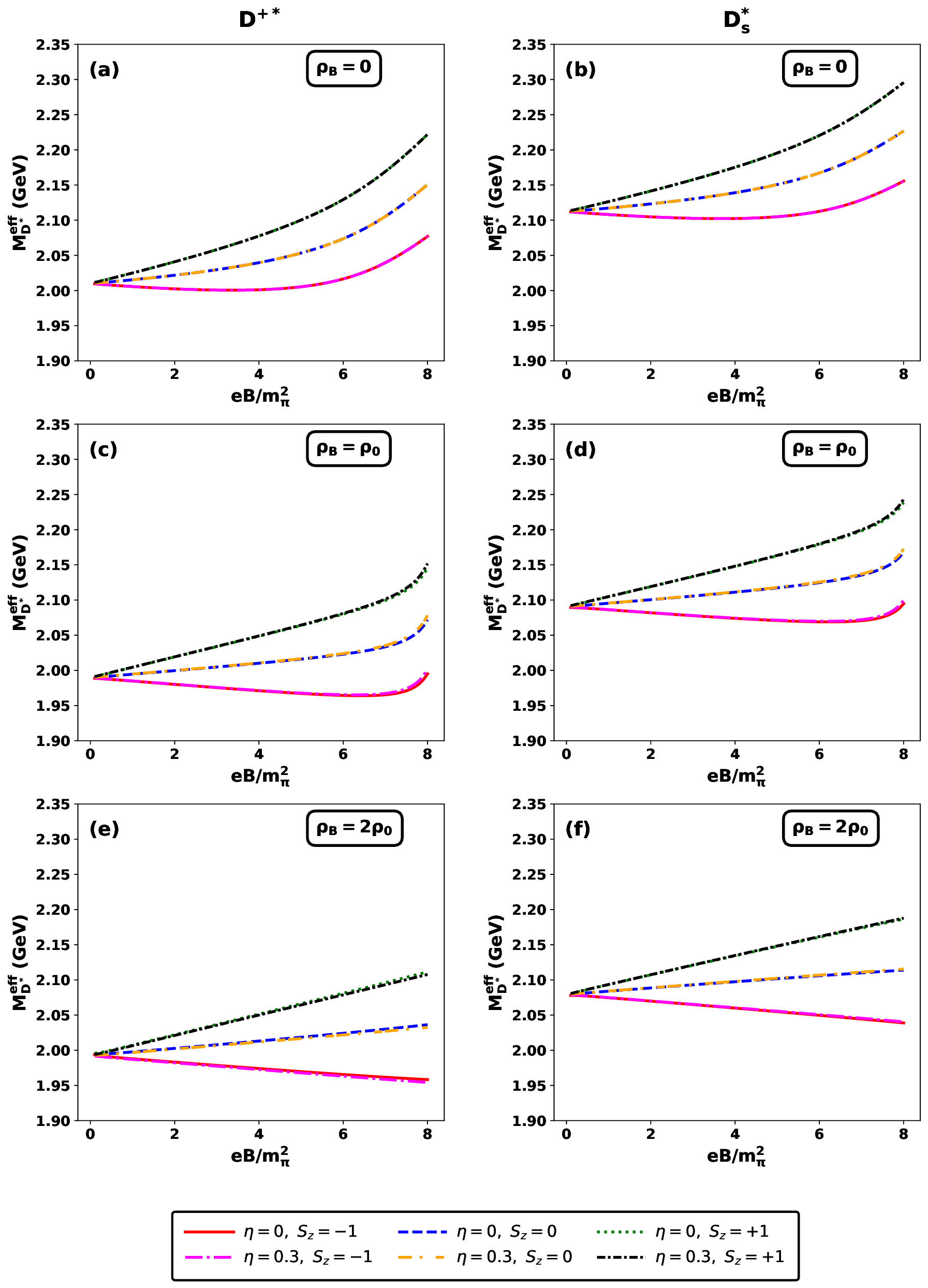}
    \caption{Effective masses of charged vector $D^{+*}$ [in subplots (a),(c),(e)] and $D_s^*$ [in subplots (b),(d),(f)] mesons as a function of magnetic field strength for all polarization states $S_z=-1,0,+1$ at fixed baryonic densities $\rho_B=0,\rho_0,2\rho_0$ in symmetric ($\eta=0$) and isospin asymmetric ($\eta=0.3$) nuclear matter at $T=0.1$ GeV.}
    \label{fig:3}
\end{figure*}

The dependence of the polarized effective masses of charged vector mesons $D^{+*}$ and $D_s^*$ on the external magnetic field is displayed in Fig.~\ref{fig:3} for different baryonic densities. The corresponding numerical values of the effective masses for different spin projections, magnetic fields, baryon densities, and isospin symmetries are listed in Tables~\ref{tab:eta0} and \ref{tab:eta0.3}. We observe that the vector mesons exhibit a different magnetic field dependence due to the coupling of their spin projections ($S_z$) with the external magnetic field through the term $gS_zeB$ in Eq.~(\ref{eq:vectorLLL}). We see that in the absence of the magnetic field, all three polarization states ($S_z=-1,0,+1$) remain degenerate because the rotational symmetry is unbroken. As the magnetic field becomes finite, this degeneracy vanishes and all three polarization state exhibit distinct masses. This observed splitting originates from the spin-dependent magnetic interaction coupled with the Landau quantization of the charged vector mesons, modifying the dispersion relation differently for each spin state. Consequently, the $S_z=+1$ state acquires the largest effective mass, while the $S_z=-1$ remains the lightest. Moreover, the separation between the polarized state increases continuously with increasing magnetic field strength, reflecting in the stronger magnetic interaction at larger values of $eB$. The baryonic medium produces an overall downward shift of the polarized branches. Since the scalar interactions with baryonic density are primarily spin independent, they lower the masses of all polarization states by almost the same amount, however, without altering their ordering. Therefore, the mass hierarchy $S_z=+1>S_z=0>S_z=-1$, established by magnetic field, is preserved throughout the entire density range. At higher densities, the attractive scalar interaction compensates the magnetic enhancement of the masses, resulting in a relatively weaker magnetic field dependence of the polarized states. We observe that the spin-dependent splitting produced by the external magnetic field follows the same behavior as that of the non-strange vector meson, indicating that the polarization splitting is controlled by the magnetic interaction rather than by the flavor composition of the meson. The effect of isospin asymmetry is found to be considerably weaker than that of the external magnetic field. However, the increase in isospin asymmetry from $\eta=0$ to $0.3$ produces a small shift in the polarized masses through the modification of the scalar-isovector $\delta$ field. 

\begin{figure*}[htbp]
    \centering
    \includegraphics[width=0.85\linewidth]{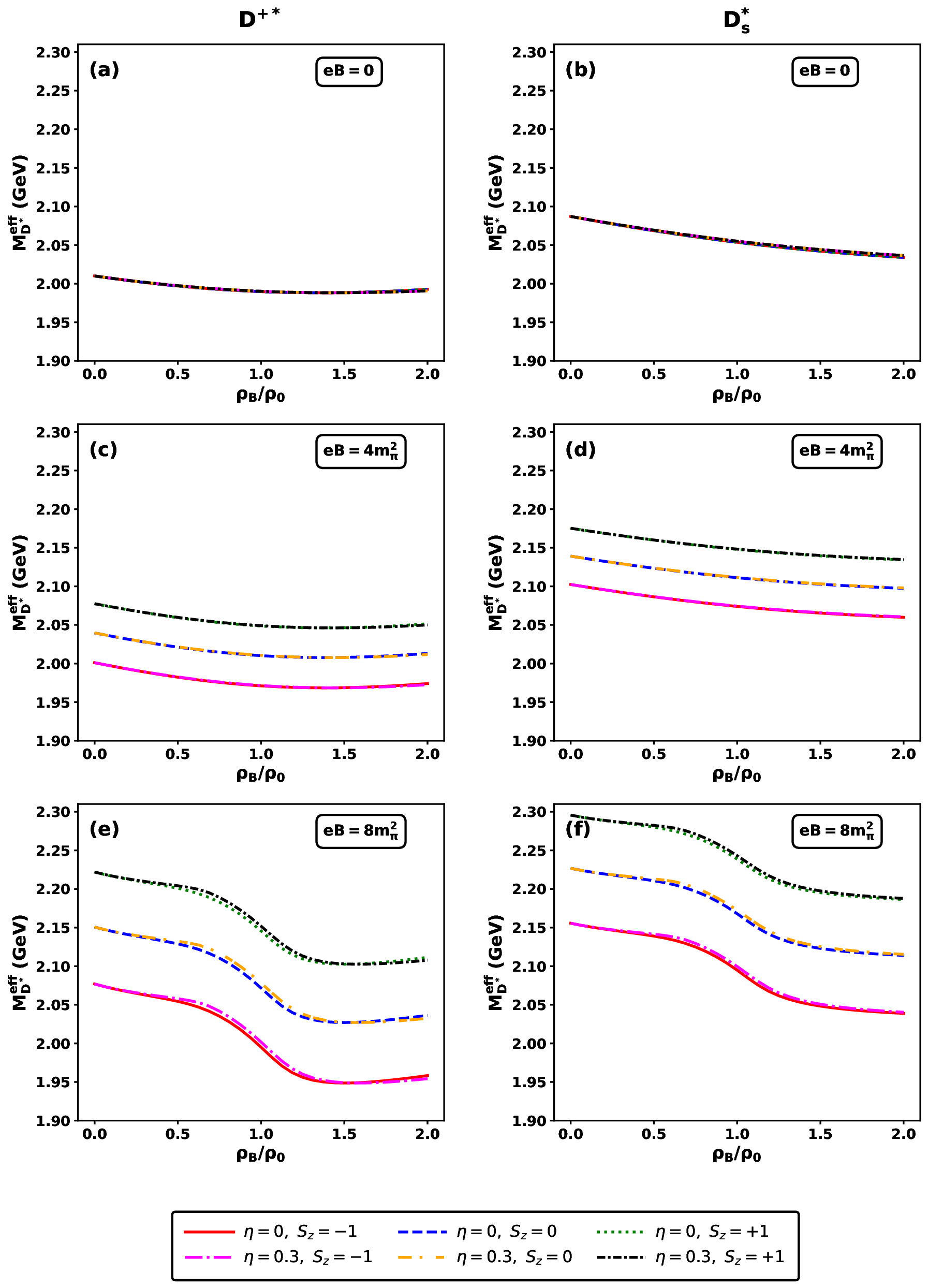}
    \caption{Variation of in-medium effective masses of charged vector $D^{+*}$ [in subplots (a),(c),(e)] and $D_s^*$ [in subplots (b),(d),(f)] mesons as a function of baryonic density for all polarization states $S_z=-1,0,+1$ at fixed magnetic field strength values $eB=0,4m_\pi^2,8m_\pi^2$ in symmetric ($\eta=0$) and isospin asymmetric ($\eta=0.3$) nuclear matter at $T=0.1$ GeV.}
    \label{fig:4}
\end{figure*}
Figure~\ref{fig:4} presents the density dependence of spin-resolved $D^{+*}$ and $D_s^*$ effective masses for different magnetic field strengths. When the magnetic field is absent $(eB=0)$, all three polarization states are degenerate and decrease with increasing baryon density. A finite magnetic field removes this degeneracy and gives rise to three distinct polarization curves whose separation becomes increasingly pronounced with increasing magnetic field strength. We see that for all magnetic field strengths, the effective masses decrease with increasing baryon density. This demonstrates that the attractive scalar interactions continue to dominate the overall medium modification even in the presence of strong magnetic fields. The reduction is particularly pronounced particularly around the nuclear saturation density. Beyond this density, the rate of decrease becomes smaller as the scalar fields gradually approach saturation, resulting in a relatively weak density dependence at higher densities. In contrast, the relative separation among the polarized states is nearly unchanged, which confirms that the magnetic field governs the spin splitting, whereas the baryonic medium determines the in-medium mass shift of the vector mesons. It is also observed that the effective masses increase slightly with increasing isospin asymmetry and effect becomes more pronounced at higher baryon densities. Since the $D^{+*}$ meson contains a light $\bar{d}$ quark, its effective mass is directly influenced by the modification of the $\bar{d}$-quark mass through $\delta$ field, compared with the $s$-quark which is relatively less sensitive to the $\delta$ field, so that the non-strange meson depends more strongly on the isospin asymmetry than the strange meson.

\begin{figure*}[htbp]
    \centering
    \includegraphics[width=1.00\linewidth]{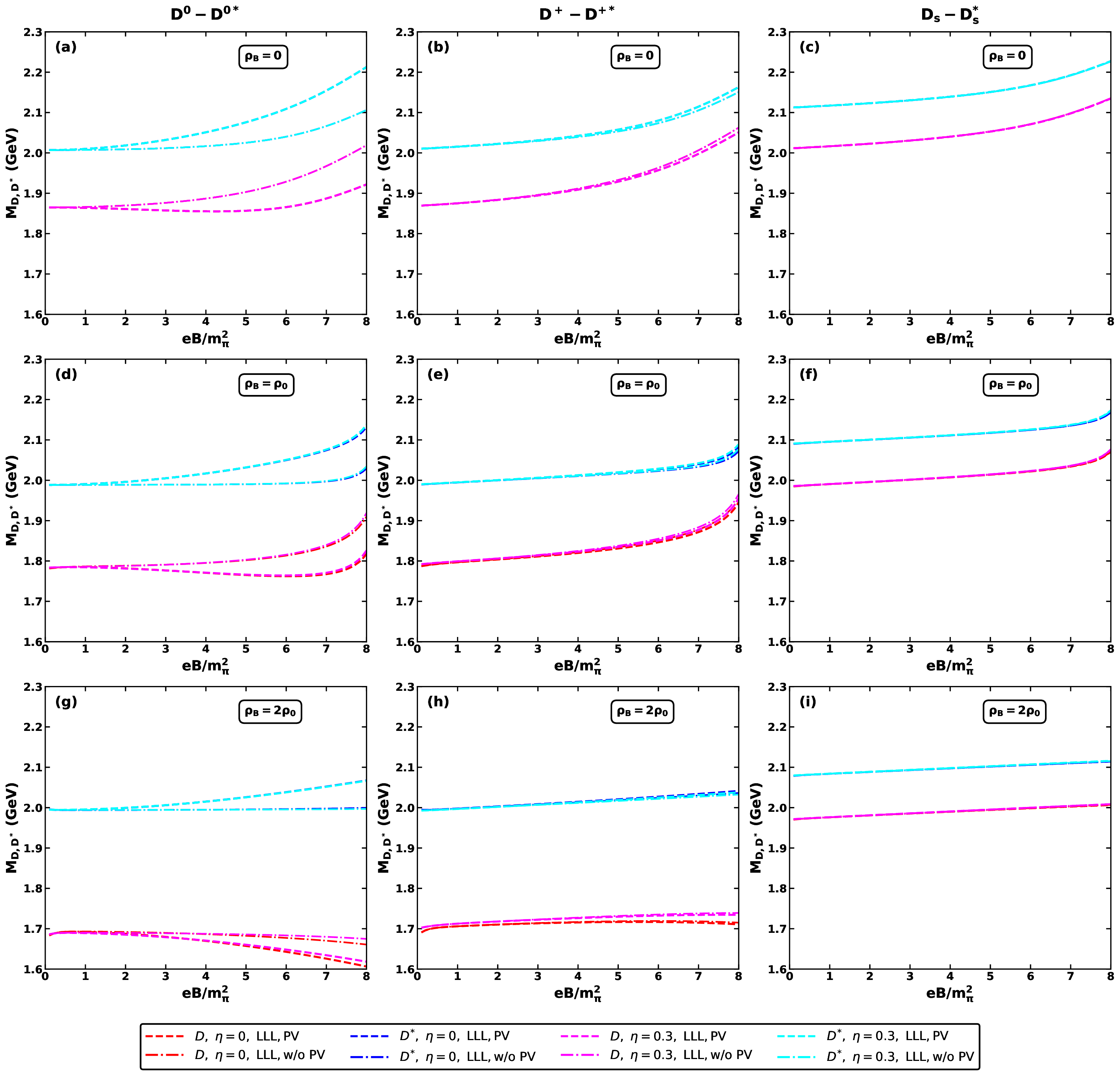}
    \caption{In-medium masses including the magnetic field induced pseudoscalar-vector (PV) mixing as a function of magnetic field strength $eB/m_\pi^2$ at different baryonic densities $\rho_B=0,\rho_0,2\rho_0$ for $D^0-D^{*}$ [in subplots (a),(d),(g)], $D^+-D^{+*}$ [in subplots (b),(e),(h)], and $D_s-D_s^*$ [in subplots (c),(f),(i)] mesons in symmetric ($\eta=0$) and isospin asymmetric ($\eta=0.3$) nuclear matter at $T=0.1$ GeV. } 
    \label{fig:5}
\end{figure*}

\begin{table*}[t]
\centering
\renewcommand{\arraystretch}{1.50}
\setlength{\tabcolsep}{10pt}

\begin{tabular}{|c|c|c|c|c|c|c|}
\hline
\multirow{2}{*}{\textbf{Meson}} &
\multicolumn{2}{c|}{$\rho_B=0$} &
\multicolumn{2}{c|}{$\rho_B=\rho_0$} &
\multicolumn{2}{c|}{$\rho_B=2\rho_0$} \\
\cline{2-7}

& $\eta=0$ & $\eta=0.3$
& $\eta=0$ & $\eta=0.3$
& $\eta=0$ & $\eta=0.3$ \\
\hline

$D^0$      & 1.9219 & 1.9219 & 1.8176 & 1.8261 & 1.6063 & 1.6183\\
\hline

$D^{0*}$   & 2.2129 & 2.2129 & 2.1318 & 2.1378 & 2.0673 & 2.0666\\
\hline

$D^+$      & 2.0508 & 2.0508 & 1.9451 & 1.9553 & 1.7108 & 1.7343 \\
\hline

$D^{+*}$   & 2.1630 & 2.1630 & 2.0824 & 2.0891 & 2.0414 & 2.0378 \\
\hline

$D_s$     & 2.1349 & 2.1349 & 2.0709 & 2.0761 & 2.0065 & 2.0084 \\
\hline

$D_s^{*}$  & 2.2274 & 2.2274 & 2.1685 & 2.1732 & 2.1145 & 2.1160 \\
\hline

\end{tabular}

\caption{In-medium masses (GeV) of pseudoscalar $D$ and vector $D^*$ mesons including the pseudoscalar-vector (PV) mixing at $eB=8m_\pi^2$ for baryon densities $\rho_B=0,\rho_0,2\rho_0$ in symmetric ($\eta=0$) and isospin asymmetric ($\eta=0.3$) nuclear matter at $T=0.1$ GeV. For the charged states, the lowest Landau level contribution is included as in Tables~\ref{tab:eta0} and \ref{tab:eta0.3}.}
\label{tab:PV_mass}

\end{table*}

Figure~\ref{fig:5} shows the effect of PV mixing on the in-medium masses of the both pseudoscalar and vector (charged and neutral) mesons. In the absence of the magnetic field, the off-diagonal mixing term vanishes and the pseudoscalar and vector mesons remain independent. As the magnetic field increases, the reduction of rotational symmetry leads to mixing of these states, producing the characteristic level repulsion between them. Consequently, the inclusion of PV mixing shifts the vector meson mass upward and the pseudoscalar mass downward as compared to the case without PV mixing. This splitting increases monotonically with $eB$, because of $(eB)^2$ term in Eq.~(\ref{eq:PVmass}). The influence is small around $eB\sim4m_\pi^2$ and becomes pronounced at larger field strengths. The corresponding numerical values of the PV-mixed masses at $eB=8m_\pi^2$ for $\rho_B=0,\rho_0,2\rho_0$ and $\eta=0,0.3$ are listed in Table~\ref{tab:PV_mass}. Comparing across the densities, however, shows that the additional impact of PV mixing on top of the LLL contribution is not uniform with density. At zero baryon density, both the PV and w/o PV curves of neutral mesons separates clearly and progressively over the full magnetic field range, while charged mesons exhibit only a negligible separation. At finite baryon density ($\rho_B=\rho_0$, subplots (d)-(f)) the splitting is reduced for both neutral and charged mesons. At $\rho_B=2\rho_0$ (subplots (g)-(i)), the neutral mesons still show a small but noticeable separation between the PV and without PV curves, especially at higher magnetic fields. In contrast, for the charged mesons both the curves nearly overlap throughout the entire $eB$ range. This is because the charged meson masses already receive a considerable contribution from the LLL through Eqs.~(\ref{eq:PS_LLL}) and (\ref{eq:vectorLLL}). Consequently, the additional mass shift arising from the pV mixing term becomes comparatively small with respect to the total effective mass. This results in an almost overlap of the curves with and without PV mixing, particularly at finite baryon densities. The dominant medium modification comes from the density-driven shift and the LLL contribution, and PV mixing acting as a smaller correction at higher-field. The magnitude of the PV mixing effect depends strongly on the corresponding radiative coupling constant $g_{PV}$, demonstrating the largest mass splitting for $D^0-D^{0*}$ and smallest for $D_s-D_s^*$ system. This indicates that the PV mixing mechanism is formally governed by the electromagnetic coupling to the internal quark-spin structure of the meson, while density-driven mass shift is governed by the scalar mean fields coupling to the light-quark mass. Furthermore, the effect of isospin asymmetry remains comparatively weak and introduces only a small shift in the masses without modifying the overall PV mixing pattern. These results show that PV mixing is different source of magnetic field induced mass modification from the Landau level and DS contributions, and directly affect the $D-D^*$ mass splitting through $D^*\to D\gamma$ decay widths in strongly magnetized nuclear matter. The behavior of open-charm mesons in strong magnetic fields has been investigated using different theoretical approaches. The present work is in good qualitative agreement with Ref.~\cite{Gubler:2015qok}, where the effects of PV mixing and Landau quantization on the masses of $D$ and $D^*$ mesons are studied within the QCD sum rule approach . It was found that PV mixing leads to a decrease in the pseudoscalar mesons masses while increasing the vector mesons masses, whereas Landau quantization provides the dominant positive contribution to the charged mesons. The in-medium masses of open-charm mesons are also investigated in magnetized nuclear matter by including the effects of Landau quantization, PV mixing, and the Dirac sea in Ref.~\cite{De:2022gse}. They observed that the in-medium masses of $D$ mesons increase with the increasing magnetic field, whereas the  baryonic density suppresses this enhancement. The present results are consistent with these studies in showing that PV mixing and Landau quantization play an important role in the magnetic-field dependence of open-charm meson masses, showing that baryon density has a much stronger effect on the meson masses than the magnetic field.

%================ DECAY CONSTANT =================%
\subsection{Decay Constants}

\begin{figure*}[htbp]
    \centering
    \includegraphics[width=1.0\linewidth]{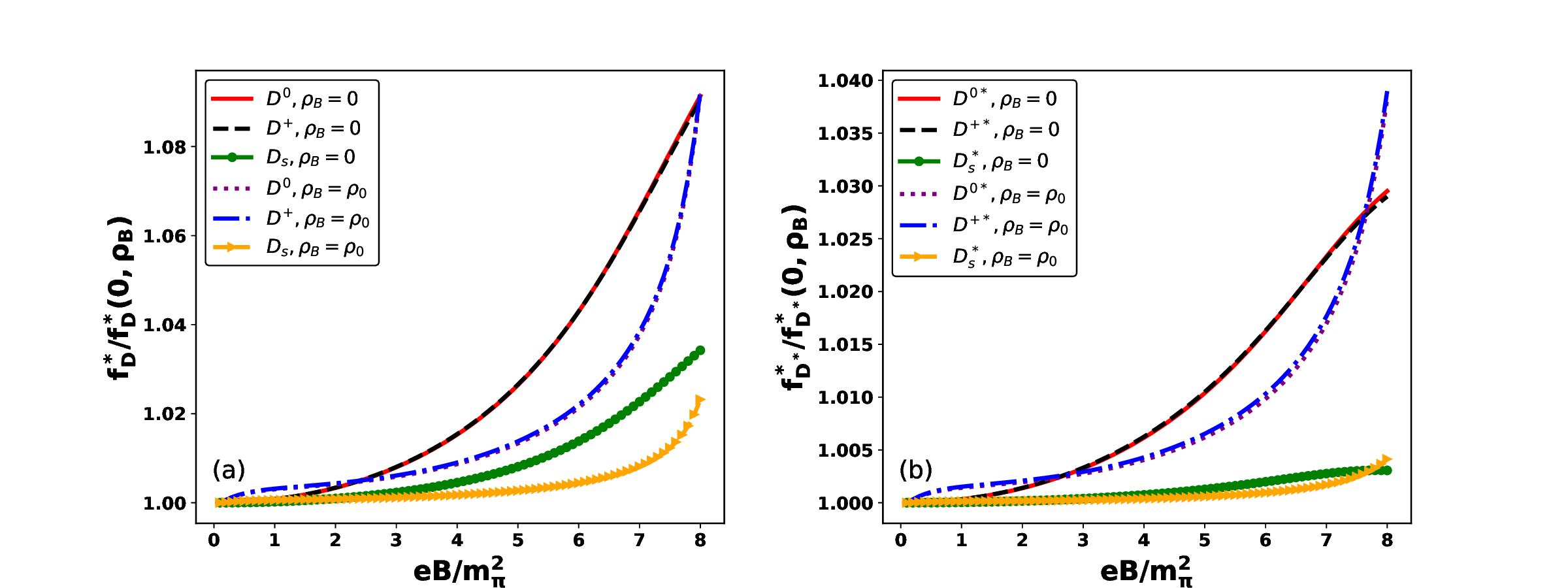}
    \caption{The in-medium weak decay constant normalized to its zero-field value at the same baryon density, $f_D^*(eB,\rho_B)/f_D^*(0,\rho_B)$, of pseudoscalar (subplot (a)) and vector (subplot (b)) $D$ mesons as a function of $eB/m_\pi^2$ at baryon densities $\rho_B=0$ and $\rho_0$ for isospin asymmetry value $\eta=0$ at $T = 0.1$ GeV.}
\label{fig:DC_eB}
\end{figure*}

Figure~\ref{fig:DC_eB}(a) and (b), respectively, show the in-medium weak decay constant normalized to its value at $eB=0$ at the same baryon density, $f_D^*(eB,\rho_B)/f_D^*(0,\rho_B)$, as a function of $eB/m_\pi^2$ at baryon densities $\rho_B=0$ and $\rho_0$ for the pseudoscalar $D$ and vector $D^*$ mesons at $\eta=0,\ T=0.1$ GeV. We see that both channels show a clear increasing trend with the magnetic field. At low values of $eB$, the increase is gradual but at high magnetic field, the enhancement becomes more pronounced. This behavior originates from magnetic catalysis: the magnetic field strengthens the light-quark condensates and increases the effective light-quark mass. As can be seen from Eq.~(\ref{eq:PS_DC}) and (\ref{eq:Vector_DC}), the weak decay constants explicitly depend on the constituent quark masses through the light-front wave function. This increase in the effective quark masses tightens the quark-antiquark binding and enhances the wave function overlap at the origin, leading to a larger decay constant. We further see that the vector mesons show relatively smaller enhancement. This is mainly due to the additional contribution associated with the transverse motion of the constituent quarks, $\frac{2\mathbf{k_\perp^2}}{\mathcal{M^*}}$ (Eq.~(\ref{eq:Vector_DC})), which leads to a different medium dependence of the vector channels. These modifications for pseudoscalar $D$ mesons have direct implications for weak decay processes and their related observables. The leptonic decay width scales as $\Gamma\left(D \rightarrow \ell \nu\right)\propto f_D^2$ \cite{Xu_2021}. This enhancement in the decay constant leads to an increase in the leptonic decay rates and thus shortens the in-medium lifetime. For $D^*$ mesons, a similar impact is observed where the enhancement in decay constant influences both weak and electromagnetic processes. The electromagnetic (dilepton) width is related to $\frac{f_V^2}{M_V}$ through vector-meson dominance \cite{Xu_2021}. Moreover, the larger decay constants for vector mesons indicate changes in their internal structure, which may influence their interaction and dissociation behavior in hot and dense matter \cite{He:2022ywp, PhysRevD.59.074015}.  \par

In symmetric nuclear medium, for magnetic field $eB=8m_\pi^2$, baryon density $\rho_B=\rho_0$, and temperature $T=0.1$ GeV, the ratios, $f_D^*/f_D^*(0,\rho_B)$ for $D^0\ (D^{0*})$ and $D^+\ (D^{+*})$ are found to be $1.0918\ (1.0382)$ and $1.0924\ (1.0389)$, respectively, while for the strange mesons $D_s\ (D_s^*)$, the values are found to be $1.0231\ (1.0041)$. This weaker response of the strange channel is because of the reduced coupling of the heavier strange quark to the scalar fields that carry the magnetic catalysis effect. The presence of finite baryon density $\rho_B=\rho_0$ leads to a systematic suppression compared to the $\rho_B=0$ case. This reduction is more pronounced for the non-strange mesons ($D^0, D^+$), which show a noticeable downward shift in their curves at finite density. Consequently, the decay constants decrease with density. This observed behavior of the weak decay constants reflects the competition between magnetic catalysis and density-induced chiral restoration.

\begin{figure*}[htbp]
    \centering
    \includegraphics[width=1.0\linewidth]{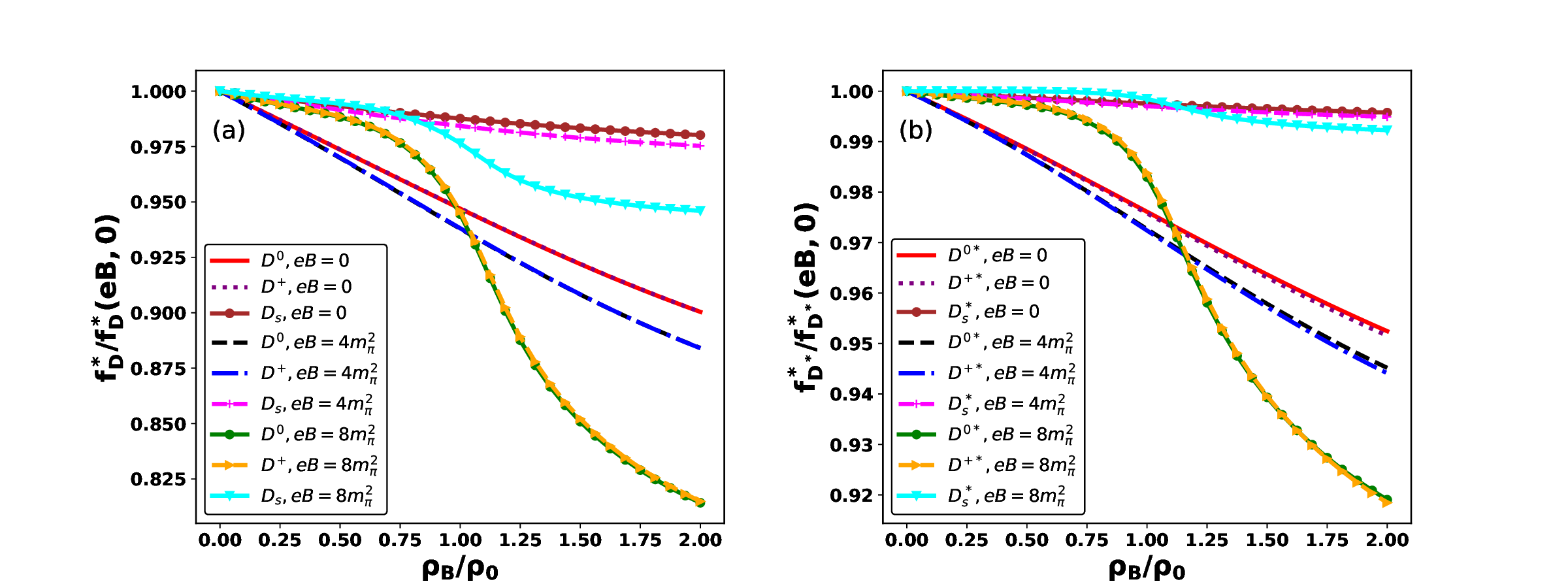}
\caption{The in-medium weak decay constant normalized to its zero-density value at the same magnetic field, $f_D^*(eB,\rho_B)/f_D^*(eB,0)$, as a function of $\rho_B/\rho_0$ for pseudoscalar (in subplot (a)) and vector (in subplot (b)) $D$ mesons at different values of magnetic field $eB$ in symmetric nuclear matter at $T = 0.1$ GeV.}
\label{fig:dc_rho}
\end{figure*}

To make picture more clean, in Fig.~\ref{fig:dc_rho} we show the density dependence of $f_D^*/f_D^*(eB,0)$, normalized at fixed magnetic field, for pseudoscalar and vector $D$ mesons at different magnetic field strengths $eB$ = 0, $4m_\pi^2$, and $8m_\pi^2$ in symmetric matter. The figure shows a clear and systematic medium modification for both pseudoscalar and vector $D$ mesons. The decay constants exhibit a decreasing trend with increasing baryonic density. At $\rho_B=2\rho_0$, $eB=8m_\pi^2$, and $T=100$ MeV, the ratios $f_D^*/f_D^*(eB,0)$ for the $D^0$ and $D^+$ mesons are found to be $0.8142$ and $0.8148$, respectively, for symmetric nuclear matter ($\eta=0$). Under the same conditions, the corresponding ratio for the $D_s$ meson is $0.9459$, indicating that the weak decay constant of the strange meson is comparatively less affected by the nuclear medium than that of the non-strange $D$ mesons. For the vector mesons, the ratios are obtained as $0.9190$, $0.9184$, and $0.9922$ for the $D^{0*}$, $D^{+*}$, and $D_s^{*}$ mesons, respectively. At zero magnetic field, the decay constants decrease gradually as the density increases, whereas at larger magnetic fields the suppression becomes much more pronounced. A clear difference is observed between the non-strange and strange mesons. The decay constants of $D^0,D^+$ mesons show a pronounced decrease with baryonic density. This stronger dependence on density can be explained by their direct coupling with the light quark condensates. On the other hand, the strange meson ($D_s$) shows only a relatively weaker density dependence, which is in agreement with the smaller medium modification of the strange quark mentioned above.  The same qualitative behavior is also observed for the corresponding vector mesons, which indicates that the mechanism responsible for the medium modification does not depend on the meson spin. However, the ratio $f_D^*/f_D^*(eB,0)$ for the vector mesons remains higher than that for the pseudoscalar mesons at all baryon densities, since the density-induced suppression is weaker in the vector channel, which clearly illustrates the role of spin-dependent interactions and medium modifications of vector condensates. The comparatively larger value for the $D_s^{*}$ meson again indicates that the strange vector meson is less sensitive to medium modifications than its non-strange counterparts.  At zero magnetic field,  the density dependence of the pseudoscalar and vector $D$ meson decay constants obtained in the present work is in good agreement with Ref.~\cite{PhysRevD.107.114010}, which reports a systematic reduction of decay constants with increasing baryonic density. Moreover, the decay constants of the $D$ mesons have also been studied in hot magnetized nuclear matter within the QCD sum rule approach by using a chiral effective model to determine the medium-modified quark and gluon condensates \cite{Kumar:2019axp}. It was found that the decay constants increase with increasing magnetic field, while finite baryon density and temperature reduce their values. The present work extends these investigations by using the CQMF-LFQM framework, taking into account the effects of the Dirac sea and the AMMs of nucleons, to examine the combined effect of the magnetic field and baryon density on the weak decay constants of both pseudoscalar and vector $D$ mesons.

%================ DA =================%
\subsection{ Distribution Amplitudes (DAs)}
\begin{figure*}[htbp]
    \centering
    \includegraphics[width=0.87\linewidth]{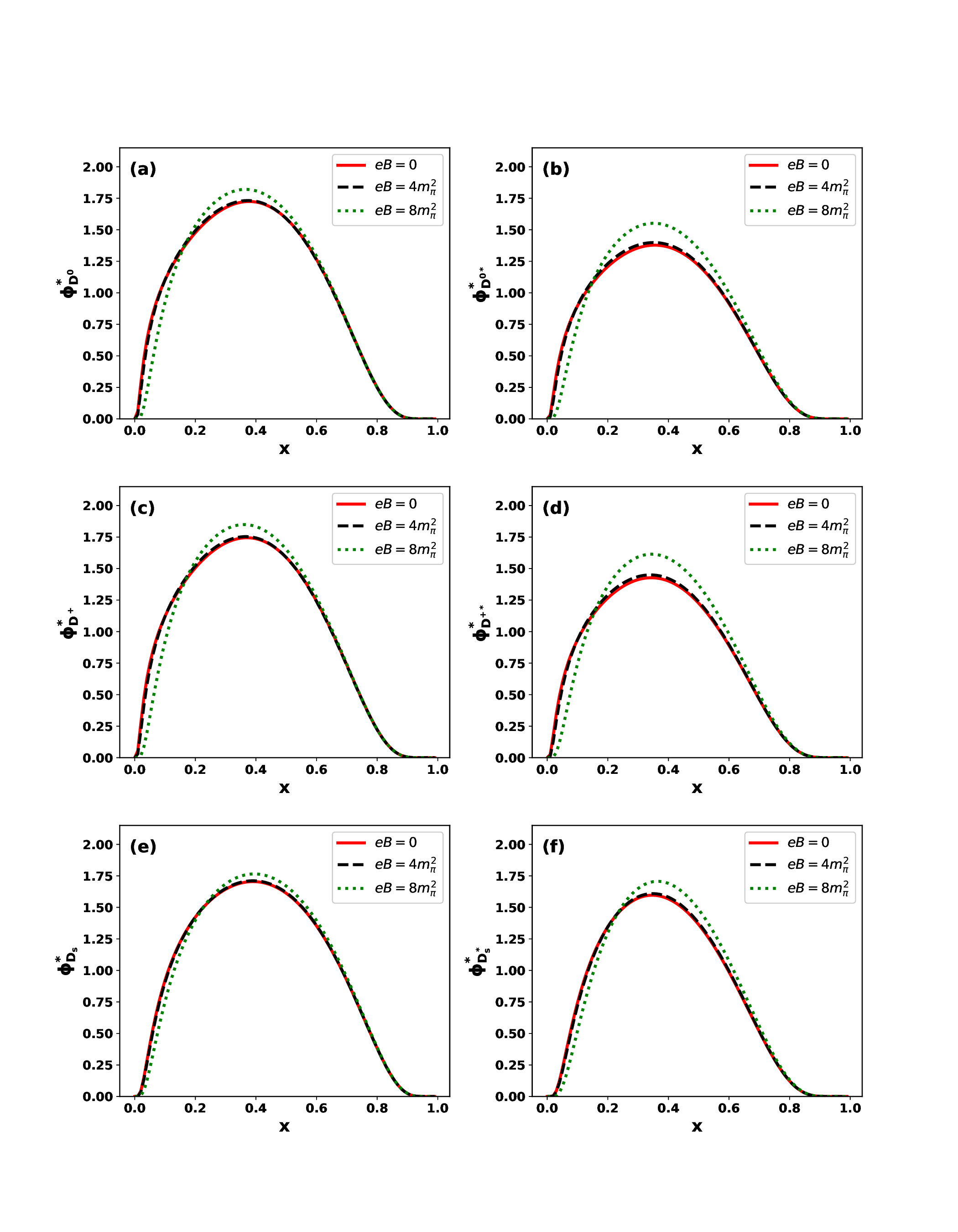}
   \caption{Distribution amplitude $\phi^*_D$ plotted for pseudoscalar [in subplots (a) $D^0$, (c) $D^+$, (e) $D_s$] and vector [in subplots (b) $D^{0*}$ , (d) $D^{+*}$, (f) $D_s^*$] $D$ mesons as a function of the longitudinal momentum fraction $x$ at different values of magnetic field $eB$ at baryon density $\rho_B=0$ and temperature $T=0.1$ GeV.}
\label{fig:DA_eB_rho_0}
\end{figure*}

Figure~\ref{fig:DA_eB_rho_0} illustrates the variation of the normalized DAs of the $D$ mesons with the longitudinal momentum fraction $x$ at different magnetic field strengths. We observe that the DA exhibits a characteristic shape for all meson species considered. In each plot, the DA curve initially rises from near zero at small $x$ and reaches a maximum at about $x=0.3-0.4$, and then decreases towards larger $x$. The peak of the DA is shifted from the symmetric point $x=0.5$ to $x\approx0.4$ which reflects the heavy-light quark structure of the $D$ mesons, where the heavy charm quark carries a larger fraction, indicating an asymmetric distribution of quarks in $D $ mesons. The inclusion of magnetic field results in a noticeable enhancement in the magnitude of the distribution around the peak region. We can see that as the magnetic field strength increases from $eB=0$ to higher values, the peak height increases systematically, indicating the strengthening of the internal quark-antiquark correlation in the presence of the magnetic field. For pseudoscalar mesons [in subplots (a),(c),(e)] and vector mesons [in subplots (b),(d),(f)], a clear distinction is observed. Both mesons follow a similar qualitative trend but the vector mesons exhibit a comparatively stronger sensitivity to the magnetic field. This is evidenced by the more pronounced separation between the curves corresponding to different $eB$ values. This enhancement effect suggests that spin-dependent interactions play an important role in determining the in-medium structure of these mesons under external magnetic field. comparing across meson types, the magnetic field does not induce uniform modifications: the strange-quark mesons display moderate changes, whereas the non-strange mesons show a larger increase in the peak values of $\phi^*_D$ which indicates a stronger response to the external magnetic field. This behavior reflects the reduced coupling of the strange quark to the surrounding medium. Overall, the results indicate that the magnetic field induces a systematic enhancement of the DAs, with the magnitude of the effect depending on both the meson type and its quark content. 

\begin{figure*}[htbp]
    \centering
    \includegraphics[width=0.85\linewidth]{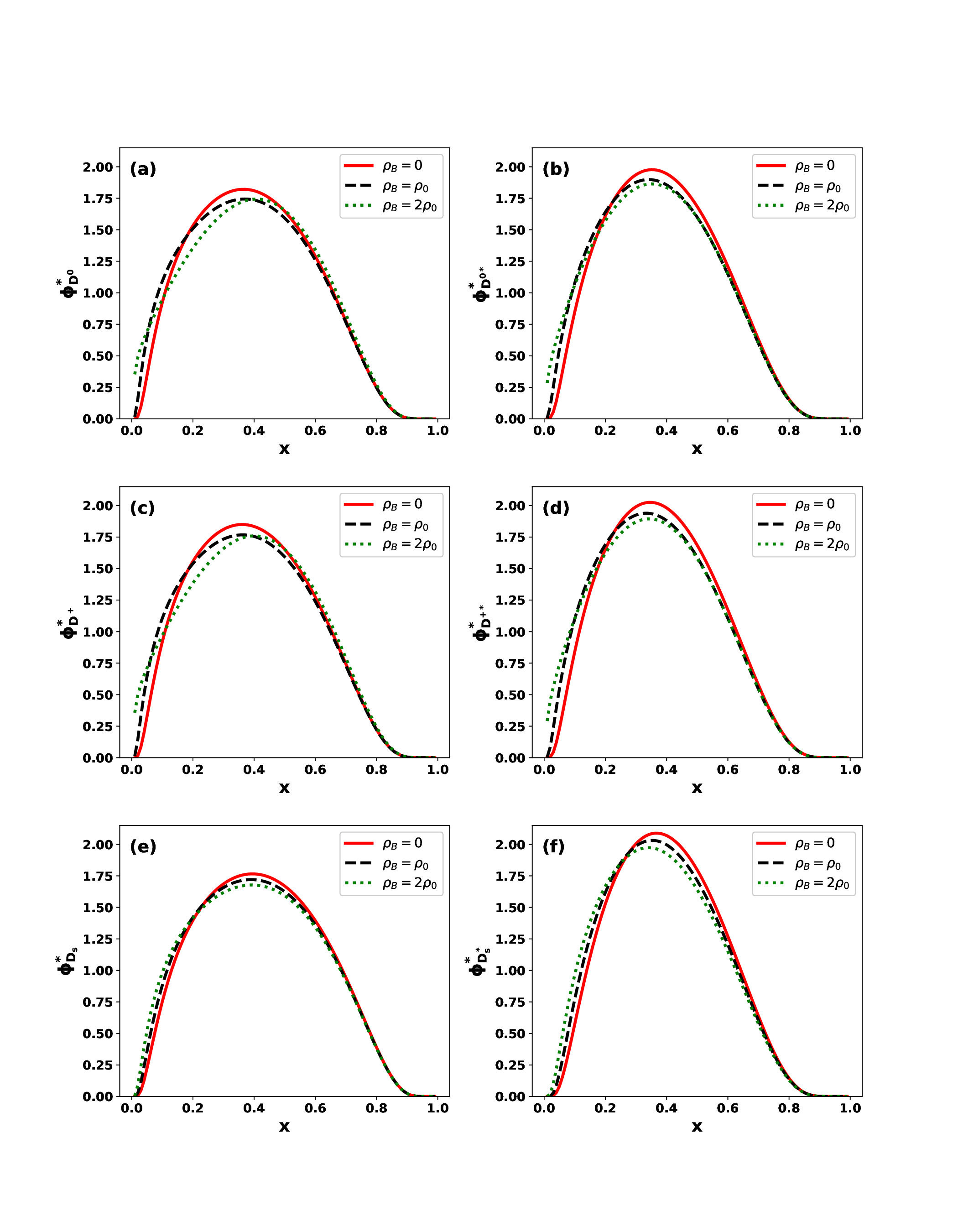}
   \caption{Distribution amplitude $\phi^*_D$ plotted for pseudoscalar [in subplots (a) $D^0$, (c) $D^+$, (e) $D_s$] and vector $D$-mesons [in subplots (b) $D^{0*}$ , (d) $D^{+*}$, (f) $D_s^*$] as a function of longitudinal momentum fraction $x$ at different values of baryonic density $\rho_B$ in the presence of external magnetic field of strength $eB=8m_\pi^2$ at temperature $T=0.1$ GeV.}
\label{fig:DA_rho}
\end{figure*}

The calculated DAs, $\phi^*_D(x)$, for the pseudoscalar and vector mesons at  different baryonic densities are presented in Fig.~\ref{fig:DA_rho} as a function of the longitudinal momentum fraction $x$. As the baryonic density increases from vacuum value $\rho_B=0$ to $\rho_B=\rho_0,2\rho_0$, the peak height decreases consistently and it causes a broadening of the DA curve. In particular, the maxima of the DA for $D^0,D^+,D_s$ are $1.7441\ (1.7436)$, $1.7675\ (1.7593)$, $1.7203\ (1.6776)$ respectively, at baryonic density $\rho_B=\rho_0\ (2\rho_0)$ at $eB=8m_\pi^2$ at temperature T=100 MeV. Similarly, for vector mesons $D^{0*},D^{+*},D_s^*$ the peaks are $1.8985\ (1.8637)$, $1.9392\ (1.8943)$, $2.0315\ (1.9738)$, respectively, under the same condition. Unlike the enhancement obtained under external magnetic field, the presence of a dense baryonic medium induces a systematic suppression in the magnitude of the DAs. The suppression becomes stronger with increasing density. We see that the strange quark containing mesons show slightly less pronounced shift compared to non-strange mesons, which may be attributed to the reduced coupling of the strange quark to the surrounding light-quark medium. \par

The DAs of both pseudoscalar and vector charmed mesons exhibit a pronounced asymmetry which reflects the unequal momentum sharing between the heavy and light constituent quarks. With increasing baryon density, the DAs undergo systematic broadening accompanied by a reduction in their peak magnitudes, indicating a weakening of the underlying quark--antiquark correlations. These medium-induced modifications are reflected in the corresponding weak decay constants and transition form factors, both of which show a decreasing trend with density. The observed behavior is consistent with the attractive in-medium mass shifts obtained for open-charm mesons. Such mass reductions lower the kinematic threshold for $D\bar{D}$ production, thereby enhancing sub-threshold open-charm yields. Consequently, these effects may influence charmonium production through the modification of decay and dissociation channels, contributing to the anomalous suppression of $J/\psi$ in dense hadronic matter \cite{MATSUI1986416}. The asymmetric nature of DAs obtained in the present work is consistent with the results of Ref.~\cite{Serna:2021xnr}, where the light-cone DAs of heavy-light mesons are studied within the Dyson-Schwinger and Bethe-Salpeter equation framework. Furthermore, the in-medium DAs of heavy-light mesons were also studied in Ref.~\cite{PhysRevD.107.114010} using a combined LFQM and quark-meson coupling model. It was found that increasing nuclear density suppresses the peak of the distribution amplitudes and broadens their shape while preserving the heavy-light asymmetry. The density-dependent behavior obtained in the present work is in qualitative agreement with these observations, whereas the inclusion of an external magnetic field leads to an enhancement of the DA peak, demonstrating the combined effect of magnetic catalysis and the dense nuclear medium on the internal momentum distribution of constituent quarks.

\section{Summary}
\label{sec:Summary}
In the present work, we have investigated the in-medium properties of pseudoscalar $D$ and $D^{*}$ mesons in magnetized nuclear matter. For this we have made use of a hybrid framework combining the chiral SU(3) quark mean field (CQMF) model and the light-front quark model (LFQM). While the medium-modified quark masses are obtained from the CQMF model, where we have included the effects of magnetized Dirac sea, the LFQM uses these masses as inputs and evaluates the in-medium meson masses, weak decay constants, and distribution amplitudes (DAs) of $D^0,D^+,D_s$ mesons and their vector counterparts. In addition to these medium effects, we have also incorporated the Landau quantization of the charged mesons $(D^{+},D_{s},D^{+*},D_{s}^{*})$ restricted to the lowest Landau level (LLL), as well as the magnetic field induced pseudoscalar-vector (PV) mixing between the pseudoscalar and the longitudinal component of the corresponding vector meson, for all three $D$-$D^{*}$ doublets. This unified approach enables a consistent description of both dynamical chiral symmetry breaking and internal quark structure of heavy-light mesons in extreme conditions such as high magnetic field and high baryon density. \par

From our results, we demonstrate that the external magnetic field leads to a significant increase in the effective masses and decay constants of the mesons, which is primarily controlled by magnetic catalysis, and which increases the effective light-quark masses and strengthens the quark-antiquark binding. In contrast, the addition of baryonic density affects the masses in the opposite way to the magnetic field which results in a reduction of the decay constant and a redistribution of the internal momentum structure. The effects of the magnetic field and baryonic density compete with each other, resulting in a nontrivial interplay in the medium: the magnetic field tends to localize and enhance the mesonic structure, whereas the dense baryonic medium induces the delocalization and suppresses central contributions. Moreover, strange mesons $D_s\ (D_s^{*})$ exhibit a weaker sensitivity due to the presence of heavier strange quark. For the charged mesons, the LLL contribution leads to a further, more pronounced enhancement of the effective mass with increasing magnetic field compared to the neutral $D^0(D^{0*})$ states. Besides the Landau and Dirac sea contributions, the PV mixing produces a level repulsion between each pseudoscalar-vector pair, which shifts the vector meson mass upward and the pseudoscalar mass downward, and this effect is strongest for $D^0(D^{0*})$ and weakest for $D_s(D_s^{*})$, indicating the dependence on the radiative coupling constant $g_{PV}$. \par

The combined effects of baryonic density and magnetic field on the masses, weak decay constants, and DAs of open-charm mesons give rise to several important phenomenological implications. The in-medium mass shifts affect the production, propagation, and interactions of open-charm mesons in dense nuclear matter. Such mass modifications are also expected to influence charmonium dissociation and regeneration mechanisms in heavy-ion collisions. The suppression (enhancement) of decay constants directly affects weak decay widths and lifetimes. On the other hand, the deformation of DAs can significantly influence hard exclusive processes and form factors in dense nuclear matter environment. The observed sensitivity of vector mesons to medium effects has measurable consequences for electromagnetic and dilepton production in heavy-ion collisions. Overall, the present study shows that a realistic description of heavy-light meson systems in extreme environments requires the inclusion of magnetic field and baryonic density effects. Present results are expected to be relevant to ongoing and upcoming experimental facilities such as FAIR, NICA and J-PARC. The study will help in understanding the behavior of hadronic matter in compact astrophysical objects.

\section*{acknowledgments}
H.D.\ would like to thank the Science and Engineering Research Board, Anusandhan-National Research Foundation (ANRF), Government of India under the SERB-POWER Fellowship scheme (Ref.\ No.\ SPF/2023/000116) for financial support. A.K.\ sincerely acknowledges Anusandhan-National Research Foundation (ANRF), Government of India for funding of the research project under the Science and Engineering Research Board-Core Research Grant (SERB-CRG) scheme (File No.\ CRG/2023/000557).

%\FloatBarrier
\clearpage
\bibliographystyle{apsrev4-2}
\bibliography{Mine}

%---------------------

 \end{document}